\documentclass[10pt]{iopart}

\usepackage{setstack}
\usepackage{graphicx}
\usepackage{amsthm}
\usepackage{amssymb}
\usepackage{eepic}
\usepackage{amscd}
\usepackage{longtable}
\usepackage{array}
\usepackage{graphicx}
\usepackage{slashbox}
\usepackage{multirow} 
\usepackage{color} 
\usepackage{bigdelim,multirow}

\newcommand{\M}{{\mathcal M}}

\newcommand{\A}{{\mathcal A}}

\newcommand{\N}{{\mathbb N}}
\newcommand{\Z}{{\mathbb Z}}
\newcommand{\C}{{\mathbb C}}
\newcommand{\one}{\mbox{1 \hspace{-3.2mm} {\bf \rm l}} }

\newcommand{\eps}{{\epsilon}}
\newcommand{\del}{{\delta}}
\newcommand{\ot}{{\otimes}}
\newcommand{\vp}{{\varphi}}

\newcommand{\frs}{{\mathfrak s}}
\newcommand{\frt}{{\mathfrak t}}
\newcommand{\fru}{{\mathfrak u}}

\begin{document}

\title{Recursive structures in the multispecies TASEP}

\author{Chikashi Arita$^{1}$,
Arvind Ayyer$^{2,3}$, \\
Kirone Mallick$^{3}$,
Sylvain Prolhac$^{4}$}

\address{$^{1}$ Faculty of Mathematics, Kyushu University, Fukuoka 819-0395, Japan\\
 $^{2}$ University of California Davis, One Shields Avenue, Davis 95616  USA\\
 $^{3}$ Institut de Physique Th\'eorique CEA, F-91191 Gif-sur-Yvette, France\\
 $^{4}$ Zentrum Mathematik, Technische Universit\"at M\"unchen, Germany}

\ead{arita@math.kyushu-u.ac.jp$^{1}$, ayyer@math.ucdavis.edu$^{2}$,
  kirone.mallick@cea.fr$^{3}$, prolhac@ma.tum.de$^{4}$}
\begin{abstract}
 We consider a multi-species generalization of
  the totally asymmetric simple exclusion process (TASEP)
  with the simple hopping rule: 
  for $\alpha$ and $\beta$th-class particles ($\alpha<\beta$), 
  the transition $\alpha\beta\to\beta\alpha$ occurs 
  with a rate independent from the 
 values $\alpha$ and $\beta$. P. A.  Ferrari and J. Martin (2007) 
  obtained the stationary state of this model thanks to  a combinatorial
  algorithm,   which was subsequently  interpreted as a matrix product
  representation by Evans {\it et al.} (2009).  This `matrix ansatz'
  shows that  the stationary state of the 
  multi-species TASEP with $N$ classes of particles ($N$-TASEP)
  can be constructed algebraically by the action of an  operator
  on  the $(N-1)$-TASEP  stationary state. 
  Besides, Arita {\it et al.} (2009) analyzed 
  the spectral structure of the Markov matrix: they showed that 
  the set of eigenvalues of the $N$-TASEP
  contains those of the $(N-1)$-TASEP and that the various spectral
  inclusions can be encoded in a hierarchical 
  set-theoretic structure  known as  the Hasse diagram.
Inspired by these works, we define  nontrivial operators
  that allow us to construct eigenvectors of the $N$-TASEP
  by lifting the eigenvectors of  the $(N-1)$-TASEP.
  This goal is achieved by  generalizing the matrix product  representation
  and the Ferrari-Martin algorithm. In particular, we show that
 the  matrix ansatz is not only a convenient 
  tool to write the stationary state
  but in fact intertwines Markov matrices of different values of $N$.
\end{abstract}

\maketitle

\section{Introduction}

Interacting particle systems are mathematical models  used 
   to study  collective properties of  N-body systems  evolving  with
   time \cite{Spohn}. 
   Replacing the actual evolution  rules (given  by a classical
   or a  quantum Hamiltonian)  by a  stochastic dynamics leads  one to
   formulate many properties of the system in terms  of  probability
   theory and allows the use of  powerful mathematical  methods
   \cite{Li85}.
   Hence, universal  properties  of  nonequilibrium statistical
   physics are  successfully  explored through  the investigation of
   stochastic  processes, that are  governed by simple dynamical rules
   at the microscopic level but that display  a rich  macroscopic
   behavior \cite{Zia,Schutz}. 
   When  defining such models,  some specific features of
   the original physical problem have to be discarded, but if the
   relevant and fundamental characteristics are retained (such as
   symmetry properties  and  conservation laws)  it is expected that
   the large scale and large time behavior of the system  will  be
   described correctly \cite{HalpinHealy}. 
   Building a simple representation for  complex
   phenomena  is  common  procedure  in  statistical  physics, leading
   to  the emergence of  paradigmatic models: the harmonic oscillator,
   the random walker, the Ising magnet and so on.
 These ``beautiful models''
   often display wonderful mathematical structures \cite{Baxter,Sutherland}.
 
In the field of nonequilibrium statistical mechanics, the
  asymmetric simple exclusion process (ASEP) has reached the  status
  of a  paradigm \cite{D98,KLS,Li99}.
   The ASEP is a lattice-gas model of interacting particles.
  Each of these particles is a random walker that hops from
  a site  to one of the neighboring  locations,  but a move is
  allowed only if the target site is empty.
This exclusion constraint mimics short-range interactions amongst particles.
   In order to drive this lattice gas out of
  equilibrium, non-vanishing  currents have to be established in the
  system and  this can be achieved by various means: by starting with a
  non-uniform initial condition that takes an infinite amount of time
  to relax, by coupling  the system to external reservoirs  that drive
  currents through the system (transport of particles, energy,
  heat, etc.) or  by introducing some intrinsic bias in the dynamics that
  favors motion in a privileged direction, each particle being  an
  asymmetric random walker that drifts steadily along the direction
  of an external driving force.
In particular, the case where moves are allowed
 in only one direction is said  totally  asymmetric (TASEP).

The ASEP has been invented several times and in different contexts
  due to its simplicity.
It was probably first proposed as a prototype
  to describe the dynamics of ribosomes along RNA \cite{MGP}.
In the mathematical literature, Brownian
  motions with hard-core interactions were
  first studied by Spitzer \cite{Spitzer}
  who coined the name  {\it exclusion process}.
The ASEP also appeared
  naturally in the description of transport processes in systems with strong
  geometric constraints such as macromolecules transiting  through
  capillary vessels and that  cannot overtake each other  \cite{Levitt}, or 
  anisotropic conductors  known  as solid electrolytes where electrons
  hop from a vacant location to another and repel each other via
  Coulomb interaction \cite{CL}.  Popular modern applications of the exclusion
  process include molecular motors that transport proteins within the
  cells along actin filaments \cite{Theo}
   and, last but not least, the ASEP and its
  variants are ubiquitous in discrete models of traffic flow \cite{Schad}.

From the mathematical point of view, the ASEP is one of the simplest
 but nontrivial models for which the hydrodynamic limit can be rigorously
 proved.
At large scales, the distribution of the particles of the ASEP
 emerges as a density field that
 evolves according to the Burgers equation with
 a vanishingly small viscosity \cite{LPS,Spohn,Varadhan}.
The Burgers equation is the textbook prototype for shock formation: 
 a smooth initial distribution can develop a singularity (a discontinuity)
 in finite time. A natural question that arises is whether  this  shock
 is an artifact of the hydrodynamic limit or if, under some
 specific conditions,  the original ASEP does display some singularity
 at the microscopic scale \cite{VanB,ABL,Cosimi,CB,JL92}. 
  This question was answered positively \cite{DJLS}:
  a shock does exist  at the  level of the particle system
  and its   width  is of the order of the
 lattice size. However, defining precisely the  position of the 
 shock at  the microscopic level requires a trick which is achieved
 by introducing a new type of particle  called a {\it  second-class particle},
 which amounts to  coupling  two TASEP models which initially differ
 at a single location \cite{Liggett76}.
  This second-class particle, denoted by  2,  has the same
 dynamics as  a normal (or first-class) particle, denoted by  1,  but
 first-class particles treat it  as a hole (denoted by 0).
The local dynamical rules thus take the
  following simple form:
\begin{equation}
 10 \to 01  \,\,\,\,  12 \to 21 \,\,\,\, \hbox{ and } \,\,\,  20 \to 02 \,,
\end{equation}
where all transitions occur with a same rate.
It was proved rigorously
 in \cite{F92,FF94} that  the dynamics of
 the  unique  second-class particle mimics the motion of the shock
 in an infinite system. Alternatively, 
  by considering a finite density of  second-class particles
  in a  periodic system,   the 
  shape  of the  microscopic   fronts were exactly calculated
  \cite{DJLS}.

A straightforward  generalization of the two species case of first- and
second-class particles is to the multispecies exclusion process  where
there is a hierarchy amongst  $N$ different species (or classes) of
particles \cite{Alcaraz,AR,MMR} ($N$-ASEP):
a particle of class $\alpha$ views  particles 
of  classes $\beta>\alpha$ as holes  and
is viewed as a hole by particles of
 classes $\beta<\alpha$;
in other words,  ``higher''  classes have lower overtaking
priority,  as will be defined  precisely  in the next section 
 \footnote{
We mention here that multispecies ASEPs with other local-interaction rules
have been introduced,
 see \cite{AHR,BE,EKKM,Karimipour}}.
 The stationary state of this simplest multispecies generalization
 of the ASEP is highly nontrivial: it was first constructed 
by Pablo Ferrari and James Martin \cite{Pablo2,FM}
in the TASEP case through a mapping to queueing processes. 
Their construction was inspired
  by earlier combinatorial results \cite{Angel,FFK94,Schaeffer}.
In  \cite{EFM,PEM}, the Ferrari-Martin algorithm was re-expressed
 as a matrix product representation in which 
the weight of a  configuration  in the $N$-ASEP 
 is written   as a linear combination of weights
of configurations of the  $(N-1)$-ASEP. 
This matrix (product) ansatz hence allows one to derive steady-state properties
 of a given  model knowing those of a {\it simpler system}.

 More general spectral
 properties of the multispecies ASEP
 on a ring were investigated in \cite{AKSS} where 
  models with different total number
 of species were related to each other in a different manner.
In that work, the key observation was that 
  particles of two consecutive classes cannot be distinguished by 
  particles of other classes.
Thus, identifying two consecutive  classes  $n$ and $n+1$
 defines a natural projection from the  $N$-ASEP  onto the  $(N-1)$-ASEP.
 Because one can identify  any two consecutive classes of particles,
 there are $N$ different such  projections.
These mappings together with 
 their commutation relations endow the set of all possible models with a 
 poset structure   represented by a Hasse diagram
 \cite{AKSS}.
 The existence of this structure leads to
 canonical inclusion relations and duality
 in the spectrum of the Markov matrix.
 In particular, through the identification mapping,
 the  eigenvectors of the  $N$-ASEP  either vanish or project
 onto eigenvectors of the  $(N-1)$-ASEP. 
  However, these  projections blur some essential information
  (namely, the difference
  between the two consecutive classes  $n$ and $n+1$)
 that cannot  be retrieved easily.
In particular, they do not allow one
 to build  the eigenvectors  of a 
 model knowing those  of a simpler system.

 The above descriptions of the matrix ansatz
 and of the  identification maps
 indicate that they operate in reverse directions and, hence, that 
 they should  be related.
This also suggests
 that the matrix ansatz may have a  range of applicability that exceeds 
 stationary-state properties;
 the matrix ansatz should be useful also to build
 ``excited states'' of the
 $N$-ASEP knowing those of the  $(N-1)$-ASEP.
In this perspective,
 the matrix ansatz could be viewed
 as a `lifting procedure' from the $(N-1)$-ASEP to the  $N$-ASEP that 
 creates a new species of particles by separating a given class into
 two consecutive subclasses. 
Further, we have seen that there are $N$ different choices for
 the identification maps from the  $N$-ASEP to 
 the $(N-1)$-ASEP.
Are there also $N$ different matrix product representations
 that would correspond to splitting a species $n$ into two
 consecutive classes $n$ and $n+1$? 
Could there be $N$ different generalizations of 
the Ferrari-Martin algorithm?

The objective of the present work is to answer these questions
 in the TASEP case
  by reformulating them in the appropriate mathematical framework
  and stating them in rigorous terms.
We will introduce a generalized matrix ansatz
which constructs a lifting operator from  the $(N-1)$-TASEP
 to the $N$-TASEP.
This will provide with a new
 and much broader perspective about this technique and shed light
 on the  recursive structures that underlie the $N$-TASEP dynamics.
As a result, the problems stated above will be solved
 and families of quadratic algebras that encode
 these  recursions will be constructed.
The outline of this work is as follows.
In section~\ref{model}, we define the dynamical rules of
 the $N$-ASEP,
 describe the characteristics and  some spectral
 properties of the Markov matrix,
 define identification operators between different
  systems (sectors) and formulate the fundamental problem addressed here.
From section~\ref{Measure} onwards,
  we restrict our consideration to the TASEP case,
  unless explicitly stated otherwise.  
In section~\ref{Measure}, we recall the Ferrari-Martin algorithm of
the stationary state of the $N$-TASEP 
 and describe the associated matrix product representation.
  The main results are derived in section~\ref{mainsection}:
  we construct a set  of generalized matrix ansatz  and we prove that
  this  allows us 
  to define a family of  conjugation operators from the $(N-1)$-TASEP
  to the  $N$-TASEP;  each of these matrix ansatz leads to  a different
  quadratic algebra 
  and we find explicit representations for all these algebras;
  finally, we find a generalized
  Ferrari-Martin algorithm corresponding to each algebra. 
  Concluding remarks are given in  section~\ref{conclusion}.
The appendices contains examples and technical details.

\section{The $N$-ASEP: Definition and  properties \label{model}}

\subsection{Definition of the model}

 The  $N$-ASEP  on the ring ${\mathbb Z}_L$ with $L$ sites
  is defined by the following dynamical rules.  Each site $i \in
  {\mathbb Z}_L$,  is assigned with a variable (local state) $k_i \in
  \{1, \ldots, N+1\}\, (N \ge 0)$.  We introduce a stochastic process
  such that  nearest neighbor pairs of local states  $(\alpha,
  \beta)=(k_i, k_{i+1})$ are interchanged with the following
  transition rates:
\begin{equation}\label{rule}
\alpha \beta \to \beta \alpha \;\;
\left\{ \begin{array}{ll} p& \mbox{if } \alpha <
\beta,\\ q& \mbox{if } \alpha > \beta,
\end{array}\right.
\end{equation}
$p$ and $q$ being  real nonnegative parameters, with the choice $q\le
p$ without loss of generality
\footnote{In the paper \cite{AKSS}, the dynamical rule is defined as
$\left\{ \begin{array}{ll} p& \mbox{if } \alpha > \beta,\\ q& \mbox{if } \alpha <
\beta.
\end{array}\right.$
We have to replace $p\leftrightarrow q$ or reverse the variables
$k_i\to N+2-k_i$, if we wish to compare the present work with
\cite{AKSS}.}. 
In particular, for $q=0$ the model is
totally asymmetric  and is called the  $N$-TASEP; 
 for $0<q<p$,  the model is said
 partially asymmetric ($N$-PASEP);  
  $q=p$ is the  symmetric case 
 ($N$-SSEP).

The dynamics is  formulated in terms of the continuous-time master
equation for  the probability $P(k_1\cdots k_L;t)$ of finding a
configuration $k_1\cdots k_L$
at time $t$:
\begin{equation}\label{a:master1}
\fl \quad
\eqalign{
\frac{d}{dt}P(k_1\cdots k_L;t) =&\sum_{i \in {\mathbb
Z}_L}\Theta(k_{i+1}-k_i) P(k_1\cdots k_{i-1}k_{i+1}k_ik_{i+2}\cdots
k_L;t) \\ &-\sum_{i \in {\mathbb Z}_L}\Theta(k_i-k_{i+1}) P(k_1\cdots
k_L;t),}
\end{equation}
where $\Theta$ is a step function defined as
\begin{eqnarray}
\Theta(x) =
\left\{ \begin{array}{ll}
 p & (x<0), \\ 0 & (x=0), \\ q & (x>0).
\end{array}\right.
\end{eqnarray}

This model can be regarded as an interacting multispecies particle
system by interpreting the local state $k_i=\alpha$ as representing
the site $i$ occupied by a particle of the $\alpha$th kind.  In the
conventional terminology, a particle of the $\alpha$th kind for
$1\le\alpha\le N$ is called ``$\alpha$th-class particle.''  We can
view the local state $N+1$ as a vacant site (or a hole), which  is
often denoted   by $0$.  However, for later  convenience, it is better
to use $N+1$ instead of $0$ and we shall stick to that convention.

\subsection{The Markov matrix}

Let $\{|1\rangle, \ldots,|N+1\rangle\}$ be the basis of the
single-site space $\C^{N+1}$.
The tensor product
  $|k_1\cdots k_L\rangle =
   |k_1\rangle\otimes\cdots\otimes|k_L\rangle\in(\C^{N+1})^{\otimes L}$
 corresponds to the configuration $ k_1\cdots k_L$
 on the ring.
The probability vector at time $t$ can be written as
\begin{eqnarray}
 |P(t)\rangle = \sum_{1\le k_i \le N+1}
 P(k_1\cdots k_L; t)|k_1 \cdots k_L\rangle .
\end{eqnarray}
 In this language, the master equation (\ref{a:master1}) becomes
\begin{eqnarray}\label{a:master2}
\frac{d}{dt}|P(t)\rangle = M^{(N)}|P(t)\rangle \, ,
\end{eqnarray}
 where the  (total)  Markov matrix   $M^{(N)}$ is   of size $(N+1)^L$
 by $(N+1)^L$. This   Markov matrix  can be written as the sum of
 local  linear operators $\left(M^{(N)}_{\rm Loc}\right)_{i, i+1}$
 acting only on  the $i$th and the $(i+1)$th components of the tensor
 product: 
\begin{eqnarray}
 M^{(N)}&= \sum_{i \in {\mathbb Z}_L} 
 \left(M^{(N)}_{\rm Loc}\right)_{i, i+1} \, ,
\end{eqnarray}
 where the   action of the local Markov matrices on a bond is given by
\begin{eqnarray}
\eqalign{
  M^{(N)}_{\rm Loc} = 
  \sum_{\alpha,\beta=1}^{N+1}
   \left(-\Theta(\alpha-\beta)|\alpha\beta\rangle\langle\alpha\beta|
      +\Theta(\alpha-\beta)|\beta\alpha\rangle\langle\alpha\beta|\right).
\label{MLoc}
}
\end{eqnarray}

Note that  off-diagonal elements of $M^{(N)}$ are $p, q$ or $0$, and
the diagonal elements are expressed as
$px+qy$ with nonpositive integers $x$ and $y$.
The sum of entries in each column of
$M^{(N)}$ is $0$, assuring the conservation of the total probability
$\sum_{1\le k_i \le N+1}P(k_1\cdots k_L; t)$.  The master equation can
be solved formally as
\begin{eqnarray}
|P(t)\rangle = \e^{t M^{(N)}}\,|P(0)\rangle,
\end{eqnarray}
and thus the eigenvalues and the right eigenvectors of $ M^{(N)}$ give
information for physical properties of the model.  The Markov matrix
has the eigenvalue 0, and we call the corresponding right  eigenvectors
stationary states.  All the other eigenvalues have  strictly negative
real parts (Perron-Frobenius theorem \cite{Spohn}),
  which characterize the relaxation to the stationary
states. We remark that  the Markov matrix $M^{(N)}$ is Hermitian only
for  $q=p$, where it becomes the  Hamiltonian of the
$sl(N+1)$-invariant Heisenberg spin chain.

\subsection{Particle conservation and  Hasse diagram structure}

In view of the transition rule (\ref{rule}), the total Markov matrix
obviously preserves  the number of particles of each kind.
Let $m_\alpha \ge 0$ denote the number of particles of the $\alpha$th class
 (respectively holes), for $1 \le \alpha \le N$ (respectively $\alpha=N+1$):
\begin{eqnarray}
\underbrace{1\cdots 1}_{m_1} \underbrace{2\cdots 2}_{m_2}\cdots
\underbrace{N+1\cdots N+1}_{m_{N+1}} .
\end{eqnarray}
Then, the state space decomposes into sectors
 labeled by  $m=(m_1,\dots,m_{N+1})$ with the constraint
 $m_1+\cdots+m_{N+1}=L$:
\begin{equation}
(\C^{N+1})^{\otimes L}= \bigoplus_{m}V_m.
\label{=otVm}
\end{equation}
The Markov matrix has  a block diagonal structure, leaving  each sector
 invariant
\begin{equation}
\label{=otMm}
M^{(N)}=  \bigoplus_{m}M_m, \quad M_m \in {\rm End} V_m,
\end{equation}
 where the square matrix $M_m$ acts on the vector space $V_m$
 spanned by  all configurations belonging to the sector $m$:
\begin{eqnarray}\label{Vm=span}
V_m=&\bigoplus_{\# \{i|k_i=j \} = m_j} \C |k_1\cdots k_L \rangle \, .
\end{eqnarray}
The dimension of  $V_m$ is given by  $\dim V_m = \frac{L!}{
 m_1!\cdots m_{N+1}!} \,$
  and the total  number of sectors for given $L$ and $N$ is  $
\left(\begin{array}{c}L+N\\ N\end{array}\right)$.

Let us call a sector $m=(m_1,\ldots, m_{N+1})$ a {\it basic sector} if
$m_n>0$ for all $n=1,\dots,N+1$
 (\textit{i.e.} there exists at least one particle of each type).
Note that there are no basic sectors for $N\ge L$.
Thus, we shall always take  $N\le L -1$.
For example, for $L=4$, the list of all basic sectors for different
 values of $N$ is given by
\begin{equation}\label{L4basicsectors}
\begin{array}{rc}
N=3:&  (1,1,1,1),\\
 N=2:& (2,1,1),(1,2,1),(1,1,2),\\
 N=1:& (3,1),(2,2),(1,3),\\
 N=0:& (4).
\end{array}
\end{equation}
The number of basic sectors for given $L$ and $N$ is given by
$\left(\begin{array}{c}L-1\\ N\end{array}\right)$.
Let $\M$ be the set of all labels for the basic sectors:
\begin{eqnarray}\label{M}
\fl \quad
\M=\{(m_1,\dots, m_{N+1}) | 0\le N\le L-1,
m_i\in\N,m_1+\cdots+m_{N+1}=L\}.
\end{eqnarray}

 We now introduce an alternative labeling of the  basic sectors
 \cite{AKSS} that will be very useful in the following:
let $s_j$ be the total number of particles of classes $k \le j$, \textit{i.e.}
\begin{eqnarray}\label{one-to-one}
s_j = m_1+ m_2 + \cdots + m_j.
\end{eqnarray}
We have $m_j = s_j-s_{j-1} > 0$ with the convention  $s_0=0$,
and thus each basic sector of the $N$-ASEP
 can be labeled by the set $\frs = \{s_1,  \dots, s_N\} $
 with $0<s_1 < s_2 < \cdots < s_N<L$.
The set $\frs$ is a subset of
 $\Omega =\{1,2,\dots, L-1\}$, \textit{i.e.}  the set $\frs$ is an
 element of ${\mathcal S}$, the power set (the set of all
 subsets) of $ \Omega.$
We can identify ${\mathcal M}$ with  ${\mathcal S}$
by the one-to-one correspondence (\ref{one-to-one}).
In the following, we shall use both labels equivalently:
 for instance, the invariant vector spaces (respectively  the  Markov
 matrices acting on them)  will be denoted either by $V_m$ or $V_\frs$
 (respectively   $M_m$ or  $M_\frs$).

\begin{figure}[h]
\begin{center}
\includegraphics[height=7cm]{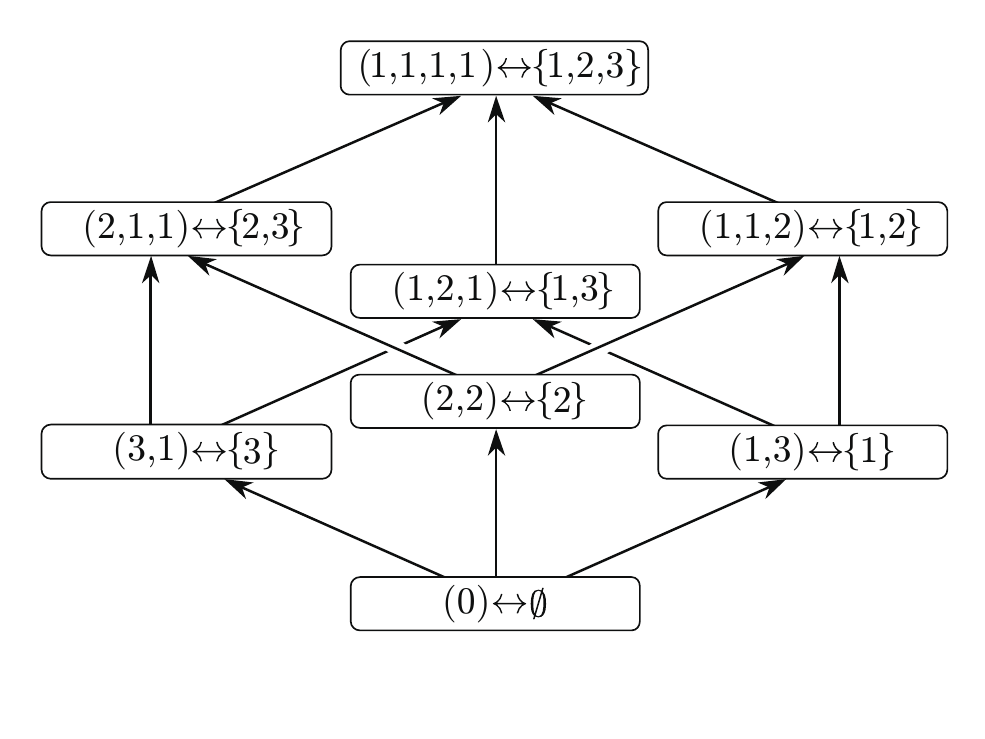}\\[-1cm]
 \caption{Basic sectors for $L=4$
in the Hasse diagram.
Each sector is labeled by an element of
$\M$ and one of $\mathcal S$.
}
 \label{Hasse}
 \end{center}
\end{figure}

 An  example of  the identification ${\mathcal M} \leftrightarrow
{\mathcal S}$ for $L=4$ is given in figure~\ref{Hasse}.
The set ${\mathcal S}$ is equipped with a natural poset (partially
ordered set) structure  with respect to the inclusion $\subseteq$,
 which is encoded in the Hasse diagram \cite{St}.
In the present case, it is just the
$L-1$ dimensional hypercube, where each vertex corresponds to a
sector.
Every link of the hypercube becomes an arrow
$\frt\rightarrow \frs$  meaning that $\frt \subset \frs$ and 
$\#\mathfrak{s} = \# \frt + 1$.
The ``maximal sector'' $\Omega
=\{1,2,\ldots, L-1\}$ corresponds to a unique sink and the ``minimal
sector'' $\emptyset$ corresponds to a unique source, as in figure~\ref{Hasse}.

The following spectral properties in
 this Hasse diagram were  proved in
\cite{AKSS}:
\begin{itemize}
\item 
\noindent
{\bf Spectral inclusion}:
Let ${\rm Spec}(\frs)$ be the multiset of the eigenvalues of $M_\frs$
with the multiplicity of an element representing the degree of its
degeneracy. Then, 
\begin{eqnarray}\label{Spec-supset-Spec}
{\rm Spec}(\mathfrak{s}) \supset {\rm Spec}(\mathfrak{t})
\end{eqnarray}
for any pair of sectors $\mathfrak{s} \supset \mathfrak{t}$.
In particular, ${\rm Spec}(\Omega)$ contains  the eigenvalues of the
Markov matrix $M_{\mathfrak s}$ of all the sectors  ${\mathfrak s} \in
{\mathcal S}$.
An example of this spectral inclusion for $L=4$ is displayed in
 figure~\ref{spec}.
\begin{figure}[h]
\begin{center}
 \includegraphics[height=9cm]{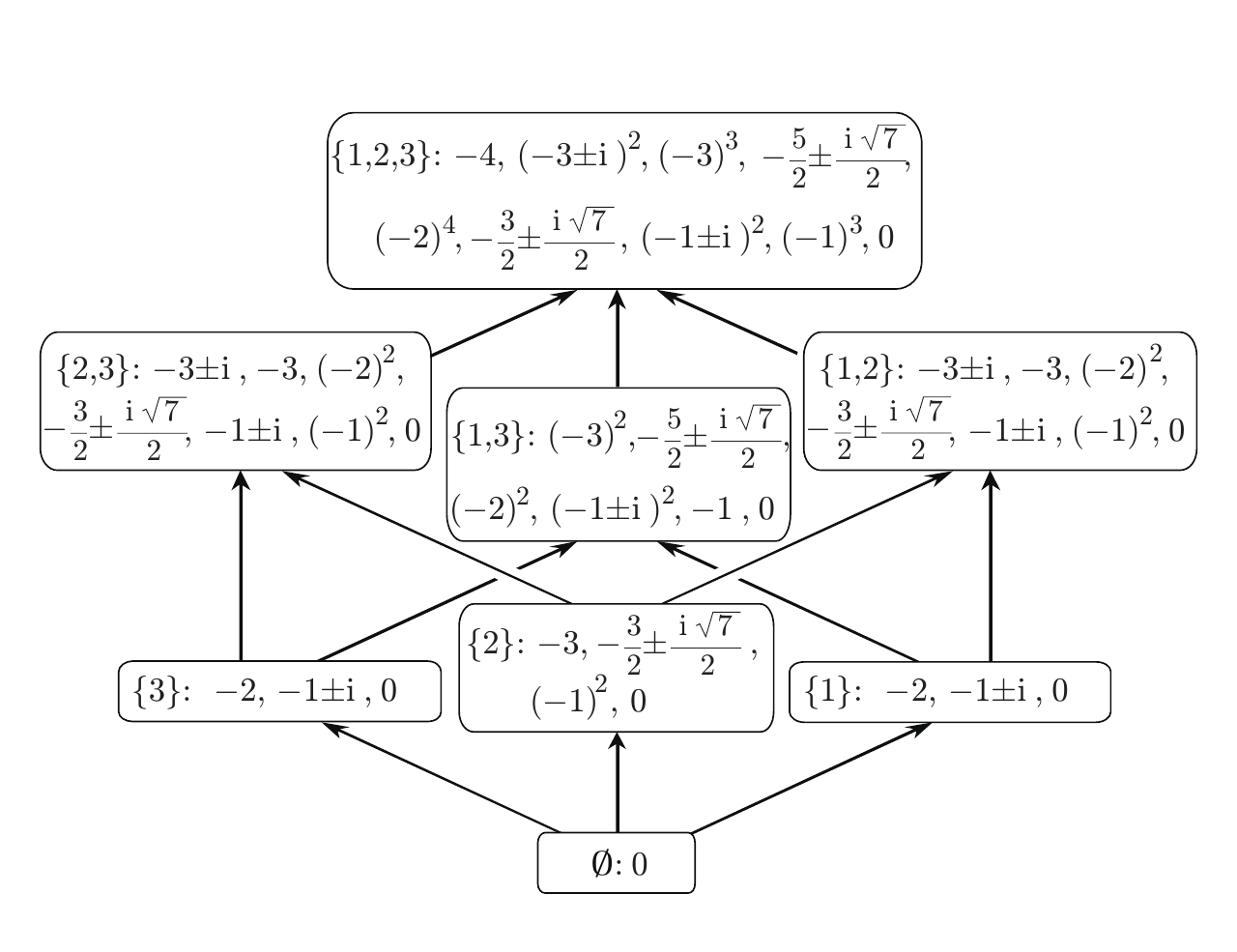} \\[-0.5cm]
\caption{${\rm Spec}({\mathfrak s})$ for $L=4$
with $p=1$ and $q=0$ (TASEP case).
Each superscript denotes the multiplicity.
For example, the Markov matrix $M_{\{1,3\}}$ has 
eigenvalue $-2$, and its degree of degeneracy is 2.}
\label{spec}
\end{center}
\end{figure}

\item
\noindent
{\bf Spectral duality}:
Using the  spectral inclusion theorem, we  can classify the eigenvalues
  $E\in\rm Spec (\frs)$ into  two types:
(i) eigenvalues that already exist in  lower sectors $\fru\subset\frs$,
(ii) eigenvalues that appear  at $\frs$.
Let us call eigenvalues of the second type {\it genuine eigenvalues}.
  More explicitly, the genuine spectrum $\rm Spec^{\circ} (\frs)$ of
  the sector $\frs$ is defined as
\begin{eqnarray}\label{inclusion}
\fl \quad
\rm Spec^{\circ} (\emptyset) :=&\rm Spec (\emptyset) =\{0\} , \quad
\rm Spec^{\circ} (\frs) := \rm Spec (\frs)
{\setminus}\bigcup_{\fru\subset\frs}\rm Spec (\fru).
\end{eqnarray}
 Then,  the following duality relation was proved:
\begin{eqnarray}\label{Spec=-L(p+q)-Spec}
{\rm Spec}^{\circ}(\mathfrak{s}) =-L(p+q) -{\rm Spec}^{\circ}
(\bar{\frs}),
\end{eqnarray}
where the sector $\bar{\frs}=\Omega\setminus\frs$ is furthest from the
sector $\frs$ in the Hasse diagram.
\end{itemize}

These two spectral relations were derived  by introducing a family of
linear operators  connecting
different sectors, that we now  review.

\subsection{The identification operators ${\varphi}_{\mathfrak{t}\mathfrak{s}}$
 and the conjugation property}

Let us consider  two (basic) sectors
$\mathfrak{s}=\{s_1<\cdots<s_N\}$ and  $\mathfrak{t} =
\mathfrak{s}\setminus\{s_{n_1}, \ldots, s_{n_u}\}$.
We introduce  a
linear operator $\varphi_{\mathfrak{ts}} : V_{\mathfrak{s}} \rightarrow V_{\mathfrak{t}}$, whose action
on the   basis  vectors is given by
\begin{eqnarray}\label{defphi}
 \fl \quad
| k_1 \cdots k_L\rangle \in V_\mathfrak{s}
 \ \mapsto\ 
| k'_1 \cdots k'_L\rangle \in V_\mathfrak{t},
\quad \mbox{with}  \ x' = x-\#\{i| n_i <x\}.
\end{eqnarray}
 Note that  $
  {\varphi}_{\mathfrak{t}\mathfrak{s}}:  V_\mathfrak{s} \rightarrow
  V_\mathfrak{t}$ is   surjective.  For  $\mathfrak{s} =
  \mathfrak{t}$,   $\varphi_{\mathfrak{s}\mathfrak{s}}$  is  the
  identity operator for any sector $\mathfrak{s}$.

To understand the general definition (\ref{defphi}),
  we give an example 
$\mathfrak{s}=\{2,3,5,8\}\supset\mathfrak{t}=\{2,5\}$  with
$L=9$.
These two sectors correspond to the following 
 compositions of the system into different types of particles:
\newcommand{\bcsp}{\!\!\!\!}
\begin{eqnarray}
\begin{array}{cccccccccccccc}
  &  &  & \bcsp\scriptstyle{2} &  & \bcsp\scriptstyle{3} &  &  & \bcsp\scriptstyle{5} &   &  & &  \bcsp\scriptstyle{8} & 
\\
\frs: &1&1& \bcsp\big| &\bcsp  2   & \bcsp\big| &\bcsp   3  &  3  & \bcsp\big| &\bcsp 4 & 4  & 4 & \bcsp\big| &\bcsp  5 
\\[1mm]
\frt: &1&1& \bcsp\big| &\bcsp  2   & \bcsp  &\bcsp   2  &  2  & \bcsp\big| &\bcsp 3 & 3  & 3 & \bcsp  &\bcsp  3 
\\
  &  &  & \bcsp\scriptstyle{2} &  & \bcsp &  &  & \bcsp\scriptstyle{5} &   &  & &  \bcsp & 
\end{array}
\end{eqnarray}
According to these lists, we define
$\varphi_{\mathfrak{t}\mathfrak{s}}$ to be the operator  replacing (or
identifying) the local states  according to the rules
   $3 \rightarrow 2, 4 \rightarrow 3, 5
\rightarrow 3$  (keeping $1$ and $2$ unchanged) within all the ket
vectors $\vert k_1\cdots k_L\rangle$ in  $V_\mathfrak{s}$.  For
example,
\begin{eqnarray}
 \varphi_{\mathfrak{t}\mathfrak{s}}|345214431\rangle
=|233213321\rangle.
\end{eqnarray}

 More generally, for a pair of sectors
     $\mathfrak{t} \subseteq \mathfrak{s}$, consider
  a chain  $\frs_0 \supset \frs_1\supset \cdots \supset \frs_u$ of
  sectors such that $\mathfrak{s}_0 = \mathfrak{s}, \mathfrak{s}_u =
  \mathfrak{t}$ and  $\# \mathfrak{s}_{j}=\# \mathfrak{s}_{j+1} + 1$
  for all $0 \le j <u$.  Then, the decomposition
\begin{eqnarray}\label{phi=phicdotsphi}
 \vp_{\frt\frs} =\vp_{\frt\frs_{u-1}}\vp_{\frs_{u-1}\frs_{u-2}} \cdots
\vp_{\frs_2\frs_1}\vp_{\frs_1\frs}
\end{eqnarray}
holds and is  independent of the choice of the intermediate sectors
\cite{AKSS}.

\subsection{The conjugation property}

A crucial property of the identification operators is that they
 provide a conjugation between different sectors.
   More precisely,  for  any   two sectors  such that   $\frs
 \supseteq \frt$  the following diagram commutes:
\newcommand{\dst}{\displaystyle}
\begin{eqnarray} \label{DiagComm}
\begin{CD}
  V_\frs  @> \dst M_\frs >>  V_\frs    \\
  @V \dst \varphi_{\frt\frs} VV  @VV \dst \varphi_{\frt\frs} V  \\
  V_\frt  @> \dst M_\frt >>  V_\frt
\end{CD}
\end{eqnarray}
which means
\begin{eqnarray}\label{Mphi=phiM}
 M_\frt \varphi_{\frt\frs} =\varphi_{\frt\frs} M_\frs \, .
\end{eqnarray}
 We shall  call this type of relation a
  {\it conjugation relation}, 
  relating the dynamics governed by $M_\frs$ and that governed by  $M_\frt$.
 The identification operator  $\varphi_{\frt\frs}$
 intertwines  the two Markov matrices
 $M_\frs$ and  $M_\frt$, and we call  $\varphi_{\frt\frs}$ a
 {\it conjugation matrix}.
 The relation (\ref{Mphi=phiM})  was the key
 \cite{AKSS} in proving
 the spectral properties (\ref{Spec-supset-Spec})
 and (\ref{Spec=-L(p+q)-Spec}).
Each right eigenvector  $|E\rangle$ with eigenvalue
  $E$ in the sector $\frs$ can be projected  down by
  ${\varphi}_{\frt\frs}$ to the sector $\frt$, and
  $\varphi_{\frt\frs}|E\rangle$ is also a right eigenvector with
  eigenvalue $E$ in the sector $\frs$ (under the assumption
  $\varphi_{\frt\frs}|E\rangle \neq 0$).  In particular,  the
  stationary state of a sector $\frs$  is mapped by the identification
  operator to the  stationary state of any sector  $\frt$ such that 
 $\frt \subseteq \frs$.

\subsection{Looking for an inverse conjugation relation}

   The identification operator  ${\varphi}_{\frt\frs}$ always maps an
 upper (more complex) sector $\frs$ into a lower (simpler) sector
 $\frt$. This implies that   identification matrix
 $\varphi_{\frt\frs}$ cannot help us to construct a right eigenvector
 in the sector  $\frs$ knowing  the  eigenvector with the same
 eigenvalue in a smaller sector  $\frt$.
It would be very useful  if
 we could {\it lift} information from lower sectors to
 upper sectors in the Hasse diagram.  One of the motivations of the
 present work can be formulated  as follows:
Can we construct a matrix $\psi_{\frs\frt}$ which lifts up right
 eigenvectors from a lower sector $\frt$ to an upper sector $\frs$?
This would be possible if we could define an operator
 $\psi_{\frs\frt}$ from  $V_\frt$ to  $V_\frs$ such that the following
 diagram commutes:
\begin{eqnarray} \label{DiagComm2}
\begin{CD}
  V_\frs  @> \dst M_\frs >>  V_\frs    \\
  @A \dst \psi_{\frs\frt} AA  @AA \dst \psi_{\frs\frt} A  \\
  V_\frt  @> \dst M_\frt >>  V_\frt
\end{CD}
\end{eqnarray}
 Equivalently, for any sectors $\frt \subseteq \frs\,$
 the conjugation matrix $\psi_{\frs\frt}$ satisfies
\begin{eqnarray}\label{psiM=Mpsi}
 \psi_{\frs\frt} M_\frt = M_\frs \psi_{\frs\frt} \, .
\end{eqnarray}
Note that the directions of the vertical arrows are opposite  as
 compared to those in the previous diagram (\ref{DiagComm}),
 which is a crucial difference.
Equivalently, comparing equation (\ref{psiM=Mpsi})
 with equation (\ref{Mphi=phiM}),
 we observe that the order of the products is opposite.
The property (\ref{psiM=Mpsi}) has the following important consequence:
Let $|E_\frt \rangle$ be any eigenvector of $M_\frt$ with eigenvalue $E$.
Then
\begin{equation}
M_\frs \left( \psi_{\frs\frt}  |E_\frt \rangle \right)  =
\psi_{\frs\frt} \left(   M_\frt  |E_\frt \rangle \right)  = E \left(
\psi_{\frs\frt}  |E_\frt \rangle \right) \, ,
\end{equation}
which means that $ \psi_{\frs\frt}  |E_\frt \rangle =|E_\frs \rangle $
   is an eigenvector of $M_\frs$  with the same  eigenvalue $E$
   (if  $\psi_{\frs\frt}  |E_\frt \rangle \neq 0$).
In other words, the map  $\psi_{\frs\frt}$ allows to lift eigenvectors from
   a smaller sector to a larger sector and provides a constructive
   information from a simpler system to a more complex one
(whereas the identification operator $\varphi_{\frt\frs}$
  erases information).
The existence of  $\psi_{\frs\frt}$ is a very nontrivial property
 of the model.
In the next section, we review
   the construction of the stationary state, which will be generalized later
   to define  $\psi_{\frs\frt}$.

\section{The stationary state of the $N$-TASEP}\label{Measure}

From this section on,
 we set $p=1$ and $q=0$, \textit{i.e.} we consider the TASEP case,
 unless explicitly stated otherwise.  In the  $N$-TASEP
 dynamics, each  particle can hop to its right nearest-neighbor site
 if the target site is empty or occupied by a higher-class particle.

A stationary state is  a right eigenvector of the Markov
 matrix corresponding to the eigenvalue 0. We denote it by $|\bar
 P\rangle = \sum_{\tau} \bar P(\tau) |\tau \rangle $:
\begin{eqnarray}\label{kernel}
 0  = M^{(N)}|\bar P\rangle \, .
\end{eqnarray}

For each sector $m$,
  the Markov matrix  has  a unique stationary state
  $|\bar P_m\rangle$
  up to an overall constant factor.
For basic sectors $\frs\leftrightarrow m$,
  we write $|\bar P_\frs \rangle = |\bar P_m
  \rangle$.  From the  ``grand-canonical'' stationary state $|\bar
  P\rangle$, which is a solution of equation (\ref{kernel}), we  extract
  the  stationary state of each sector, by restricting the components
  of  $|\bar P\rangle$ to that sector.
 We emphasize that 
the multispecies TASEP on a ring does not satisfy
  the detailed-balance condition and exhibits 
   non-vanishing currents in its  stationary state,
  which  is one of the major characteristics 
  of nonequilibrium systems\footnote{ However, it is important to 
   keep in mind  that 
  boundary conditions are  absolutely  crucial 
 in  nonequilibrium physics: indeed, in a closed
 segment with  {\it reflecting} boundaries,
 the detailed-balance condition  {\it   is}  satisfied  \cite{BE}}.

The stationary state of the $N$-TASEP is  nontrivial as
  soon as $N \ge 2$.
The stationary state of the  $2$-TASEP was
  constructed by using  a matrix product representation in
  \cite{DJLS}.  However, this technique did not seem easily
  generalizable to higher values of $N$ (see \cite{MMR} for an attempt
  for the case  $N=3$).
The solution to this problem came from
  two completely different directions.
In \cite{Angel} it was shown that
  the stationary state provided by the matrix  product representation for the
  $2$-TASEP could be interpreted in terms of weights in a binary tree.
On the other hand, in \cite{FFK94}, these weights were rewritten  in terms of
  Dyck paths which also appear as trajectories of a queueing
  process and therefore the  $2$-TASEP  was  reinterpreted  as  a
  queueing process.  This fact   was generalized to the  $N$-TASEP
   which was mapped into a   system of coupled  queueing processes
  \cite{FM}:  this  construction leads 
  to  the Ferrari-Martin algorithm for  the  stationary state of
  the  $N$-TASEP that we review in the next subsection.

\subsection{Review of Ferrari and Martin's construction}

\begin{figure}[h]
\begin{center}
 \includegraphics[height=8cm]{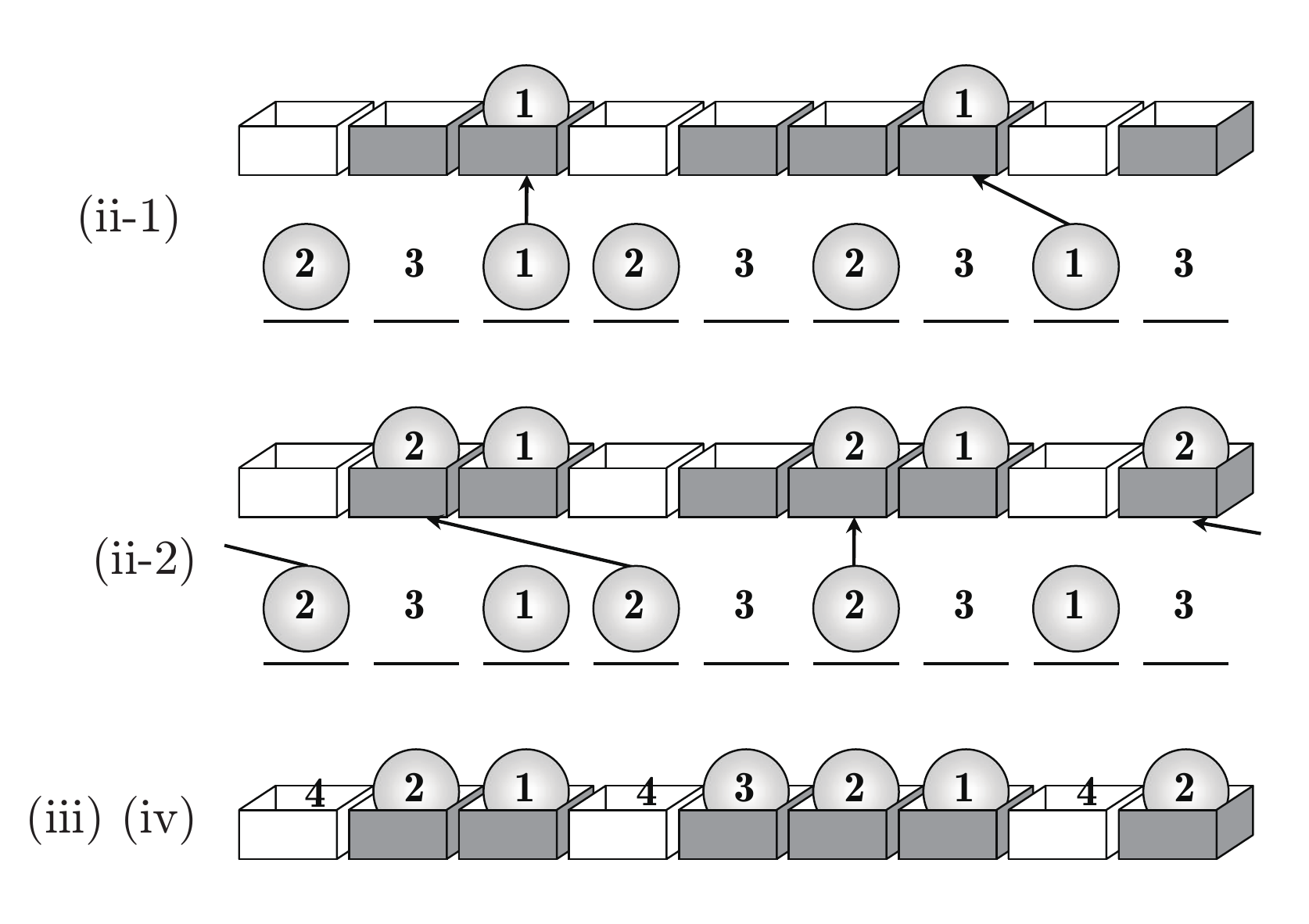}
  \caption{The Ferrari-Martin algorithm that constructs
 a configuration of $N$-TASEP from
 that of $(N-1)$-TASEP.
 This figure provides the specific  example: 
 $F(wbbwbbbwb$,231232313)$=$421432142.
 }
 \label{FM}
\end{center}
\end{figure}

We now reformulate the algorithm found in \cite{FM}
  that constructs the stationary state of the  $N$-TASEP,
  in terms of  the convention adopted in the present work.
The basic idea is to obtain the stationary state of $N$-TASEP
  from  that of $(N-1)$-TASEP.
The following algorithm is valid for any
  sector $\frs=\{s_1<\cdots<s_N\}$;  In  figure~\ref{FM}, we provide
  an explicit example for  a ring of size  $L=9$ and with
  $\frs=\{2,5,6\}$ (\textit{i.e.} $m_1 =2, m_2 =3, m_3=1$
   and $m_4 =3$). 

In order to relate a configuration
 of the   $(N-1)$-TASEP to a  configuration of  the  $N$-TASEP,  we  consider two lines, each
  of them  corresponding to a   lattice of size $L$. 
\begin{itemize}
\item[(i)] On the  upper line, we 
set $s_N$ black boxes $b$  and $(L-s_N)$ white boxes $w$
  arbitrarily, that we write  as $c_1\cdots c_L$ (with $c_i=b,w$).
\hfill\break
  On the  lower line, 
  there is  a configuration $k_1\cdots k_L$ of the 
  $(N-1)$-TASEP corresponding to
  the sector $\frs\setminus\{s_N\}$: thus,  on the  lower line, 
  there are ($s_\nu-s_{\nu-1}$) $\nu$th-class particles
 ($1\le \nu \le N-1,s_0=0$)
 and ($L-s_{N-1}$) $N$th-class particles (\textit{i.e.} empty sites).  
\item[(ii-1)] 
Let $\{ i_1^{(1)}, \dots, i_{s_1}^{(1)}\}$
  be the positions of the $s_1$ first-class particles on the lower line. 
For the first first-class particle located at $i_1^{(1)},$ 
   find the nearest black box $c_{i'}=b$ with $i'\le i_1^{(1)}$
   and put a  particle of type  1 on it.
If there is no such black box,
  put the particle 1 on the rightmost black box.
For the second first-class particle located at $i_2^{(1)},$
find the nearest unoccupied black box $c_{i'}=b$ with $i'\le i_2^{(1)}$
   and put a  particle  of type 1 on it.
If there is no such black box,
  put the particle  of type  1 on the rightmost unoccupied black box.
Iterate this procedure $s_1$ times: \textit{i.e.}, 
find the nearest unoccupied black box $c_{i'}=b$ with $i'\le i_\ell^{(1)}$
   for the $\ell$th first-class  particle  located at  $i_\ell^{(1)},$
   and put a  particle  of type  1 on it
  or on the rightmost unoccupied black box
  if  $i'$ does not exist.
\item[(ii-2)]
Now we consider the  second-class particles on the lower line.
 Recall that there are $m_2 = s_2-s_1$ of them, and set
 their positions as
 $\{ i_1^{(2)}, \dots, i_{s_2-s_1}^{(2)}\}$.
There are $(s_{N}-s_1)$ unoccupied black boxes remaining on the upper line.
 We must  iterate the following procedure $(s_2-s_1)$ times:
find the nearest unoccupied black box $c_{i'}=b$ with $i'\le i_\ell^{(2)}$
   for the $\ell$th  second-class particle  ($1 \le \ell \le s_2-s_1$),
   and put a  particle of type 2 on it
  or on the rightmost unoccupied black box
  if  $i'$ does not exist.
\item[(ii-$\nu$)] In the same way, we go on for
  third-, fourth-, $\cdots$, $(N-1)$th-class particles:
 for the  $\nu$th-class particles, 
 there are $(s_\nu-s_{\nu-1})$ $\nu$th-class particles on the lower line,
 with  positions  $\{ i_1^{(\nu)}, \dots, i_{s_\nu-s_{\nu-1} }^{(\nu)}\}$.
There are $(s_N-s_{\nu-1})$ unoccupied black boxes remaining on the upper line.
Iterate the following procedure $(s_\nu-s_{\nu-1})$ times: 
find the nearest unoccupied black box $c_{i'}=b$ with $i'\le i_\ell^{(\nu)}$
  for the $\ell$th  $\nu$th-class particle ($1 \le \ell \le s_{\nu}-s_{\nu-1}$), 
  and put a  particle of type $\nu$ on it
  or on the rightmost unoccupied black box
  if  $i'$ does not exist.
\item[(iii)] Now, there are $(s_N-s_{N-1})$ unoccupied black boxes that remain.
Put particles of type $N$ on them.
\item[(iv)]  Regarding the $(L - s_N)$ white boxes as an empty sites,
 \textit{i.e.} as particles of type  $N+1$, we have thus constructed
 a  well-defined configuration 
  $F(c_1\cdots c_L, k_1\cdots k_L)$ of the $N$-TASEP on the upper line, belonging
 to the sector  $\frs$, starting from  a configuration  $k_1\cdots k_L$
 of the  $(N-1)$-TASEP on the lower  line that was in the sector 
  $\frs\setminus\{s_N\}$.  (Note that an $N$ represents an empty site
  in the sector $\frs\setminus\{s_N\}$
  whereas  $N+1$ does in $\frs$.)
\end{itemize}

Note that {\it different} configurations of 
$(N-1)$-TASEP on the lower line
can lead to a {\it same} configuration 
of the $N$-TASEP on the upper line:
for example,
\begin{eqnarray}
 F(bbwwb,32133) =
 F(bbwwb,31323) = 21443.
\end{eqnarray}

 It was proved in \cite{FM}
  that the stationary weight of a given  configuration  in  the sector $\frs$
 is given (up to normalization) by the sum of the weights of all 
   configurations in $\frs\setminus\{s_N\}$ that are related to it through this
 construction. Equivalently, we have 
\begin{eqnarray}
  \left|\bar P_{\frs}\right\rangle
  =\sum
   |F(c_1\cdots c_L, k_1\cdots k_L)\rangle
   \langle k_1\cdots k_L|\bar P_{\frs\setminus\{s_N\}} \rangle .
\end{eqnarray}
Here the summation $\sum$ runs over
  $ c_1\cdots c_L $ and $k_1\cdots k_L$
  with $\#\{i|c_i=b\}=s_N$ and
  $k_1\cdots k_L$ belonging to the sector $\frs\setminus\{s_N\}$.

\subsection{Matrix product representation for the stationary state}
\label{reviewpem} 

The Ferrari-Martin algorithm was restated as 
  a matrix product representation in \cite{EFM}.
The basic idea of the matrix product representation
is to express the  stationary probability
 as the trace of a product of matrices
 over a suitable algebra.
This technique, initially invented
 in \cite{DEHP} for  the (one-species) ASEP
 with open boundaries,
 has been generalized to many stochastic interacting particle systems
  including discrete-time updates, 
  a second-nearest neighbor interaction
  and non-conservative dynamics (see \cite{BE} for an exhaustive
  review).
  For the $N$-TASEP on $\Z_L$, the stationary weight of 
  a configuration $j_1 \dots j_L$ is given 
by the trace of an $L$-fold matrix product:
\begin{eqnarray}\label{mpr}
\bar P(j_1\cdots j_L)=
{\rm Tr}\left[X^{(N)}_{j_1} \cdots X^{(N)}_{j_L} \right],
\end{eqnarray}
or equivalently as
\begin{eqnarray}\label{mprvec}
|\bar P \rangle= 
{\rm Tr}
\left(\begin{array}{c}X^{(N)}_1 \\ 
      \vdots \\ X^{(N)}_{N+1} \end{array} \right)^{\ot L}.
\end{eqnarray}

In \cite{PEM}, an explicit solution for the operators
 $X^{(N)}_J$'s was found. It is given by the following 
 tensor product recursions 
\begin{eqnarray}\label{X=sum_aX}
X^{(N)}_J=\sum_{K=1}^{N}a^{(N,N)}_{JK}\otimes X^{(N-1)}_K
 \quad \hbox{for  } 1 \le J \le N+1 \, , 
\end{eqnarray}
with $Y\ot X_1^{(1)}=Y\ot X_2^{(1)}=Y$ ({\it i.e.}, 
 $ X_1^{(1)} =  X_2^{(1)}=  1).$
The operators $a^{(N,N)}_{JK}$ are given in the following table:
\begin{eqnarray}
\fl \quad
\label{AlgebrePEM}
\begin{array}{|c|c|c|}
\hline 
\ & \ & \  \\[-8pt]
{}_{\displaystyle J}\diagdown K
    & 1\quad \cdots\quad N-1 & N 
\\[2pt] \hline \ & \ & \   \\[-8pt]
  1\ \cdots\  N-1
 & \begin{array}{c} A ^{\ot (J-1)} \ot\del \ot \one^{\ot(K-J-1)} \\
    \ot \eps  \ot \one^{\ot(N-K-1)}
    \end{array}
& A  ^{\ot(J-1)}  \ot \del  \ot \one^{\ot(N-J-1)}
\\[2pt] \hline \ & \ & \  \\[-8pt]
 N & 0 & A ^{\ot (N-1)}
\\[2pt] \hline \ & \ & \   \\[-8pt]
 N+1
 & \one^{\ot(K-1)} \ot\eps  \ot\one^{\ot(N-K-1)} 
 & \one^{\ot (N-1)}
\\[2pt] \hline
\end{array}.
\end{eqnarray}
We read {\rm $\del \ot\one^{\ot(-1)}\ot\eps =\one$}
and {\rm $\del\ot\one^{\ot x}\ot\eps =0$ for $x\le -2$}.
The  operators   $\del,\eps$ and $A$   are the fundamental
 building blocks that are ubiquitous in the matrix ansatz technique
 \cite{BE}. These three  operators  generate a quadratic algebra
 and satisfy the following relations: 
\begin{eqnarray}
 \label{q=0-algebra} 
  \del\eps=\one,\quad \del A=0,\quad A\eps=0.
\end{eqnarray}
A common representation of this algebra is given by
 the infinite dimensional matrices
\begin{eqnarray}
\fl
\eqalign{
 \del =
\left(\begin{array}{ccccc}
0  &   1   &      &       &   \\
   &   0   &  1   &      &   \\
   &      &   0  &   \!\! 1 &    \\[-2mm]
   &      &      &   \!\! 0  & \!\! \ddots  \\[-2mm]
   &      &      &      &  \!\! \ddots
\end{array}\!\!\!\right),\ 
 \eps=
\left(\begin{array}{ccccc}
0  &  &      &       &   \\
1  &   0   & &      &   \\
  &  1 &   0  &  &   \\
  &      & 1  & \!\! 0  & \\[-2mm]
  &      &      & \!\!\ddots &   \!\!\ddots
\end{array}\!\!\!\right),\ 
A=\left(\begin{array}{ccccc}
1  &  &      &       &   \\
&  0 & &      &   \\
  &  &   0 &  &   \\
  &      &  & 0 & \\[-2mm]
  &      &      &  &   \!\!\ddots
\end{array}\!\!\!\right) .
}
\end{eqnarray}

\hfill\break
{\bf Remark:}
 The Ferrari-Martin algorithm can not be easily 
 defined for  the PASEP because
 the directionality plays a crucial role in the algorithm. Nevertheless,
 the matrix product representation can readily be  generalized
 to the PASEP case ($p=1$ and $q \neq 0$)
 as follows: in the table (\ref{AlgebrePEM}), we 
  replace the operators $\del,\eps$ and $A$ by 
 the `$q$-deformed' operators $\del_q,\eps_q$ and $A_q$   that 
   generate a quadratic algebra with  the following relations
\begin{eqnarray}
\del_q \eps_q -q\eps_q \del_q = (1-q)\one,\quad
\del_q A_q = qA\del_q,\quad
A_q \eps_q =q\eps_q A_q.
\label{q-algebra}
\end{eqnarray}
Then, the matrix product representation (\ref{mprvec})
  with this deformation provides
  the stationary state of the $N$-PASEP, 
  as was shown in \cite{PEM}.

\subsection{Interpretation of the matrix ansatz as a linear mapping}

 The recursion relation (\ref{X=sum_aX}) implies that 
 each stationary weight of the $N$-ASEP 
    can be expressed as a linear combination
  of weights of the $(N-1)$-ASEP
  \cite{PEM}. More precisely, using
 equation (\ref{X=sum_aX})   we can write
\begin{eqnarray}
  |\bar P^{(N)}\rangle=\Psi^{(N,N)} |\bar P^{(N-1)}\rangle,
\end{eqnarray}
where the matrix $\Psi^{(N,N)}$ is defined in terms of its elements as
\begin{eqnarray}
  \langle j_1\cdots j_L|\Psi^{(N,N)}| k_1\cdots k_L\rangle
  =\Tr\left[a^{(N)}_{j_1k_1}\cdots a^{(N)}_{j_Lk_L}\right]
\end{eqnarray}
for $1\le j_i \le  N+1$ and $1\le k_i \le  N$.
Moreover the matrix $\Psi^{(N,N)}$ has 
 the following property that we shall call {\it sector specificity}:  
 suppose that  the configuration $j_1\cdots j_L$ belongs to
  a  basic sector $\frs=\{s_1<\cdots<s_N\}$ and  that the  element 
$\langle j_1\cdots j_L|\Psi^{(N,N)}| k_1\cdots k_L\rangle$
  is nonzero, then  the configuration $k_1\cdots k_L$
  belongs to the sector $\frs\setminus\{s_N\}$.
Conversely, if $j_1\cdots j_L$ belongs to $\frs$
  and $k_1\cdots k_L$ does not belong to $\frs\setminus\{s_N\}$, then 
   $\langle j_1\cdots j_L|\Psi^{(N,N)}| k_1\cdots k_L\rangle = 0$. 
We will prove this property  in appendix B.
 This property allows us to consider  
  the mapping ${\psi}_{\frs,\frs\setminus\{s_N\}}$: 
 $ V_{\frs\setminus\{s_N\}} \to  V_\frs  $,   defined as the restriction 
 of  $\Psi^{(N,N)}$ to the sectors
 $\frs$ and $\frs\setminus\{s_N\}$. 
 This mapping provides us with  a  construction of   the stationary state of
 the  basic sector $\frs$  by lifting up that of $\frs\setminus\{s_N\}$:
\begin{eqnarray}\label{bP=spibP}
|\bar P_{\frs}\rangle =
\psi_{\frs,\frs\setminus\{s_N\}}
|\bar P_{\frs\setminus\{s_N\}}\rangle \, . 
\end{eqnarray}
Using this equation repeatedly, we have
\begin{eqnarray}\label{oneway}
|\bar P_{\frs}\rangle =
\psi_{\frs,\frs\setminus\{s_N\}}
\psi_{\frs\setminus\{s_N\},\frs\setminus\{s_{N-1},s_N\}}
\cdots
\psi_{\{s_1,s_2\}\{s_1\}}
\psi_{\{s_1\}\emptyset}|1\cdots 1\rangle ,
\end{eqnarray}
 where $|1\cdots 1\rangle$ is the only configuration of the
 0-TASEP on a ring of size $L$ (the configuration in which all
 sites are empty). Hence, 
the stationary state of the  basic sector $\frs = \{s_1,\dots,s_N\}$
  is constructed along the way
\begin{eqnarray}
  \emptyset\to\{s_1\}\to\{s_1,s_2\}\to\cdots
  \to\frs\setminus\{s_{N-1},s_N\}\to\frs\setminus\{s_N\}
  \to\frs
\end{eqnarray}
in the Hasse diagram.

On the other hand, using (\ref{phi=phicdotsphi}), 
  we can project  $|\bar P_\frs\rangle$ down to the minimal sector
  via arbitrary intermediate sectors, \textit{i.e.} 
  for any $\{n_1,\dots,n_N\}=\{1,\dots,N\}$,
\begin{eqnarray}\label{phicodtsphiP=111}
  \vp_{\emptyset\frs_{N-1}}
  \vp_{\frs_{N-1}\frs_{N-2}}
    \cdots
  \vp_{\frs_{2}\frs_{1}}
  \vp_{\frs_{1}\frs}
  |\bar P_{\frs}\rangle
= \mbox{constant} |1\cdots 1\rangle ,
\end{eqnarray}
where $\frs_x=\frs\setminus\{ s_{n_1},\cdots, s_{n_x}\}$.
Comparing equations (\ref{oneway}) and (\ref{phicodtsphiP=111}),
 we observe that the   ${\psi}$ mappings  play
  a role  opposite to that of the  $\vp$'s. The  ${\psi}$'s  are thus
 good candidates to be solutions
  to the conjugation relation (\ref{psiM=Mpsi}). 
 This property will be proved in the next section.

 We emphasize  that the matrix ansatz that we have
 considered  above  allows us to construct the stationary state  only
 along  a very specific path in the Hasse diagram: in 
 the sector  $\frs\setminus\{s_N\}$, there are $m_1 = s_1$
 particles of class~1, $m_2 = s_2 -s_1$ particles of class~2,...,
 $m_{N-1} = s_{N-1} - s_{N-2} $
 and  $m_N = L -s_{N-1}$ particles of class~$N$. In the sector 
 $\frs$, the number $m_j$  of particles of class~$j$, with $ 1 \le j \le N-1$
 is the same, but there are new $m'_N = s_{N} -s_{N-1}$ particles
 of class $N$ and  $m'_{N+1} = L -s_{N}$ particles of class $N+1$,
 with  $m'_{N+1}+ m'_N = m_N$.
 Hence, when  the sector  $\frs\setminus\{s_N\}$
 is lifted up  to  $\frs$, a new species is created by splitting
 the particles of  class  $N$ (that have the  lowest priority)  into two 
 subspecies of  class  $N$ and $N+1$.  In the Ferrari-Martin algorithm,
 this means that {\it  a new species is created from the holes.}  However,
  in the Hasse diagram, there exist  different paths between two 
 connected but non-adjacent sectors.
 This observation  suggests that the matrix ansatz
  and the Ferrari-Martin algorithm should  be generalized in order 
  to construct the stationary state via arbitrary intermediate sectors.

In the next section, we show that the matrix ansatz not only
 provides a way to write the stationary state  but  also  allows one
 to define mappings that  satisfy
 the conjugation relation (\ref{psiM=Mpsi}) \footnote{
This interpretation of the  matrix ansatz
 as an intertwining operator between different dynamics was already used
 in a  recent study of a  one-species TASEP with open boundaries
 and  with annihilation \cite{AM}.
There,  conjugation matrices with respect
 to the system size were constructed,
 which allowed them to calculate 
 the normalization factor  and certain correlation functions (see 
  \cite{Woelki} for a related approach).}.
Therefore, this technique provides a tool  to lift eigenvectors by
intertwining the dynamics corresponding to  different values of $N$.
 Besides, we shall also construct intertwining operators between 
 arbitrary sectors, by defining and using more general 
 quadratic algebras than those considered  previously.

\section{Conjugation matrices from a generalized matrix ansatz}
\label{mainsection}

In this section, we  derive a general conjugation relation
 between $N$-TASEP models with different values of $N$ by using
 a generalized  matrix ansatz which allows us to create a new class of 
 of particles by splitting any intermediate species into two subspecies.

\subsection{Generalized quadratic algebra}

We define a  family of rectangular matrices 
  $\left\{a^{(N,n)}\right\}_{1\le n\le N}$
  of size $(N+1) \times N$, indexed by the integer  $n$.
The elements $a^{(N,n)}_{JK}=\langle J | a^{(N,n)} |K\rangle $
  of the matrix  $a^{(N,n)}$ are  operators given in the following table:
\begin{eqnarray}
\fl\quad
\begin{array}{|c|c|c|c|}
\hline 
\ & \ & \ & \  \\[-8pt]
{}_{\displaystyle J}\diagdown K
    & 1\quad \cdots\quad n-1 & n& n+1\quad \cdots\quad N
\\[2pt] \hline
 \begin{array}{cc}1\\[-3pt] \vdots\\[-3pt] n-1 \end{array}
 \bcsp&\bcsp
  \begin{array}{cc} A^{\ot (J-1)}\ot\del\ot\one^{\ot(K-J-1)} \\
                  \ot\eps\ot\one^{\ot(N-K-1)} \end{array}
 \bcsp&\bcsp
  \begin{array}{cc} A^{\ot(J-1)} \ot \\ \del \ot \one^{\ot(N-J-1)} \end{array}
 \bcsp&\bcsp
  \begin{array}{cc} A^{\ot(J-1)} \ot\del\ot\one^{\ot(K-J-2)} \\
    \ot\del\ot \one^{\ot(N-K)} \end{array}
\\[2pt] \hline \ & \ & \ & \  \\[-8pt]
 n \bcsp&\bcsp 0 \bcsp&\bcsp
  \begin{array}{cc} A^{\ot (n-1)} \ot \\ \one^{\ot(N-n)} \end{array}
 \bcsp&\bcsp
   \begin{array}{cc} A^{\ot(n-1)} \ot\one^{\ot(K-n-1)} \\
    \ot\del \ot \one^{\ot(N-K)}\end{array}
\\[2pt] \hline \ & \ & \ & \  \\[-8pt]
 n+1
 \bcsp&\bcsp
  \begin{array}{cc} \one^{\ot(K-1)}\ot\eps\ot \\
    \one^{\ot(n-K-1)} \ot A^{\ot(N-n)} \end{array}
 \bcsp&\bcsp
    \begin{array}{cc} \one^{\ot (n-1)}  \ot \\ A^{\ot(N-n)} \end{array}
 \bcsp&\bcsp 0
\\[2pt] \hline
\begin{array}{cc}n+2\\[-3pt] \vdots\\[-3pt] N+1 \end{array}
 \bcsp&\bcsp
  \begin{array}{cc} \one^{\ot (K-1)} \ot\eps\ot\one^{\ot(J-K-3)} \\
                  \ot\eps\ot A^{\ot(N-J+1)} \end{array}
 \bcsp&\bcsp
  \begin{array}{cc} \one^{\ot(J-3)}\ot \\ \eps\ot A^{\ot(N-J+1)}  \end{array}
 \bcsp&\bcsp
  \begin{array}{cc} \one^{\ot(K-2)} \ot\del\ot\one^{\ot(J-K-2)} \\
    \ot\eps\ot A^{\ot(N-J+1)} \end{array}
\\[2pt] \hline
\end{array}
\label{solution}
\end{eqnarray}
where we read {\rm $\del\ot\one^{\ot(-1)}\ot\eps=1$}
and {\rm $\del\ot\one^{\ot x}\ot\eps=0$ for $x\le -2$}.
 The fundamental operators   $\del$, $\eps$ and $A$  satisfy the relations
 given in  equation (\ref{q=0-algebra}). 
 In general, each element of $a^{(N,n)}$
   is  either 0 or an $(N-1)$-fold tensor product of
   $\one, A, \del$ or $\eps$.
 Some  examples are given in appendix A.
Note that for the case $n=N$, one retrieves the operators
 that were given in the table (\ref{AlgebrePEM}).

Let us define 
\begin{equation}
 \Psi^{(N,n)}=\Tr\left[\left(a^{(N,n)}\right)^{\ot L}\right] \,.
 \label{def:Psi}
\end{equation}
We shall show that the following relation is satisfied:
\begin{eqnarray}
  M^{(N)}\Psi^{(N,n)} = \Psi^{(N,n)} M^{(N-1)} \, .
\label{fondConj}
\end{eqnarray}
 In other words, for any value of $n$, {\it the matrix  $\Psi^{(N,n)}$
 allows us to embed the system  with
 $(N-1)$ classes of particles into the system
 with $N$ classes of particles.} 
 Using  $\Psi^{(N,n)}$,  we shall be able to construct 
 sector specific  conjugation operators that intertwine  the  dynamics 
 between any two  basic sectors  along {\it any
 path }  in  the Hasse diagram. 

We now derive  equation (\ref{fondConj}).
The method 
  used is an extension of the hat matrix
  technique, that was developed 
  to prove  various matrix product representations \cite{HPS,BE}.
Suppose that, for each
 value of $n$,  there exists an operator valued $(N+1)\times N$ matrix
   $\widehat{a}^{(N,n)}$ such that the following identity,
 that we shall call the {\it hat relation}, is satisfied
\begin{eqnarray}
\eqalign{
\label{hatrelation}
M^{(N)}_{\rm Loc}(a^{(N,n)}\otimes a^{(N,n)})-
(a^{(N,n)}\otimes a^{(N,n)})M^{(N-1)}_{\rm Loc}
\\
= a^{(N,n)} \otimes\widehat{a}^{(N,n)}-\widehat{a}^{(N,n)}\otimes a^{(N,n)} \, .
}
\end{eqnarray}
  Then,   equation (\ref{fondConj}) is a consequence of this relation.
   Indeed, from 
 the relation (\ref{hatrelation}), we obtain, taking the $L$-fold tensor product,
\begin{eqnarray}
\eqalign{
&\sum_{i\in\Z_L}\left(M^{(N)}_{\rm Loc}\right)_{i,i+1}
\left(a^{(N,n)}\right)^{\ot L}-\left(a^{(N,n)}\right)^{\ot L}
\sum_{i\in\Z_L}\left(M^{(N-1)}_{\rm Loc}\right)_{i,i+1}  \\
=&\sum_{i\in\Z_L}\left(a^{(N,n)}\right)^{\ot i}
   \ot\widehat a^{(N,n)}\ot\left(a^{(N,n)}\right)^{\ot(L-i-1)} \\
 &- \sum_{i\in\Z_L}\left(a^{(N,n)}\right)^{\ot(i-1)}
   \ot\widehat a^{(N,n)}\ot\left(a^{(N,n)}\right)^{\ot(L-i)}
 = 0.
}
\end{eqnarray}
 Taking the trace of this relation on the space on which the
 operators $a^{(N,n)}$ act,
 and noting that local Markov matrices sum up to  total Markov matrices,
 we obtain equation (\ref{fondConj}).
We emphasize
 that the hat matrices $\widehat{a}^{(N,n)}$ are  used  in the proof
 but  do not appear in the final result  (\ref{fondConj}). 

 To summarize, the conjugation  relation (\ref{fondConj}) follows
 from the hat relation (\ref{hatrelation}) and in order to show that
 this latter relation exists, we need to specify 
 the operators  $\widehat{a}^{(N,n)}$.
 We claim that the   hat relation (\ref{hatrelation})
 is satisfied for the choice 
\begin{eqnarray}\label{defofhat}
\widehat a^{(N,n)}=d_n a^{(N,n)} \quad \hbox{ with } \quad 
d_n={\rm diag}(\underbrace{1,\cdots,1}_{n},\underbrace{0,\cdots,0}_{N+1-n}).
 \end{eqnarray}
This explicit  expression  of $\widehat{a}^{(N,n)}$ 
  leads to closed  quadratic relations  between the elements
  $a_{JK}=\langle J | a^{(N,n)} |K\rangle$\  $(1\le J\le N+1,1\le K\le N) $
 (Note that for simplicity, we write $a_{JK}$ instead of  $a^{(N,n)}_{JK}$). 
Indeed, using the expressions  of the local Markov matrices, we have 
\begin{eqnarray}
\fl
  \langle JJ'| M^{(N)}_{\rm Loc} \left(a^{(N,n)}\ot a^{(N,n)}\right) |KK'\rangle =
  \left\{ \begin{array}{ll}
     - a_{JK}a_{J'K'} & (J<J'),  \\
                   0 & (J=J'), \\
      a_{J'K}a_{JK'} & (J>J'),  
  \end{array}\right.  \label{LHShat1}
\\
\fl
  \langle JJ'| \left(a^{(N,n)}\ot a^{(N,n)}\right) M^{(N-1)}_{\rm Loc} |KK'\rangle =
  \left\{ \begin{array}{ll}
    -a_{JK}a_{J'K'} + a_{JK'}a_{J'K} & (K<K'), \\
      0  & (K\ge K').
  \end{array}\right.  \label{LHShat2}
\end{eqnarray}
Besides, using equation (\ref{defofhat}), we  can calculate
  each element of the right hand side of (\ref{hatrelation}) as 
\begin{eqnarray} \label{RHShat}
\fl
  \langle JJ'| \left(a^{(N,n)}\ot \widehat{a}^{(N,n)}
        -  \widehat{a}^{(N,n)}\ot a^{(N,n)} \right)|KK'\rangle
 =
  \left\{ \begin{array}{ll}
     -a_{JK}a_{J'K'} & (J\le n<J'),  \\
      a_{JK}a_{J'K'} & (J>n\ge J'), \\
     0 & (\rm otherwise).
  \end{array}\right.
\end{eqnarray}

Substituting equations (\ref{LHShat1}), (\ref{LHShat2})
 and (\ref{RHShat})
 into the hat equation (\ref{hatrelation}) leads to the following 
 closed quadratic algebra generated by the operators $a_{JK}$:
\begin{eqnarray}\label{algebra}
\fl\quad
\begin{array}{|l|c|c|}
\hline 
\ & \ & \  \\[-8pt]
  &  {\rm I}\ (K<K') & {\rm II}\ (K\ge K') 
\\[2pt] \hline \ & \ & \  \\[-8pt]
{\rm A}\ (J\le n< J')
 &  a_{JK}a_{J'K'} = a_{JK'}a_{J'K}
 &  -
\\[2pt] \hline \ & \ & \  \\[-8pt]
 {\rm B}\ \left(
  \begin{array}{l} J<J'\le n \\
 {\rm or}\ n<J<J'\end{array}\right)
 &  a_{JK'}a_{J'K}=0
 &  a_{JK}a_{J'K'}=0
\\[2pt] \hline \ & \ & \  \\[-8pt]
 {\rm C}\ (J=J')
  & a_{JK}a_{JK'}= a_{JK'}a_{JK}
  & -
\\[2pt] \hline \ & \ & \  \\[-8pt]
{\rm D}\ (J>n\ge J')
 & a_{J'K}a_{JK'}= a_{JK'}a_{J'K}
 & a_{JK}a_{J'K'} = a_{J'K}a_{JK'}
\\[2pt] \hline \ & \ & \  \\[-8pt]
 {\rm E}\ \left(
\begin{array}{l} J>J'>n \\ {\rm or}\ n\ge J>J'\end{array}
\right)
 & \begin{array}{l} a_{JK}a_{J'K'}+a_{J'K}a_{JK'}
   \\  = a_{JK'}a_{J'K}  \end{array}
 & a_{J'K}a_{JK'}=0
\\[2pt] \hline
\end{array}
\end{eqnarray}
(Note that
the cases A-II and C-II
do not give any nontrivial relation, and
E-II for $K>K'$ is equivalent to B-I.)
Showing that the matrix $\Psi^{(N,n)}$(\ref{def:Psi}) satisfies
the conjugation relation (\ref{fondConj})
therefore reduces to checking
that $a^{(N,n)}_{JK}$'s defined by the table (\ref{solution})
actually give a representation for (\ref{algebra}).
Thus the proof of the conjugation relation (\ref{fondConj})
   reduces to a purely mechanical procedure.
We have checked this for several values of $(N,n)$
by using Mathematica.

 We emphasize that the key ingredient is the generalized
  hat relation (\ref{hatrelation}) together with the
  ansatz (\ref{defofhat}) for the hat matrices $\widehat a^{(N,n)}$
  which allows one to define a quadratic algebra.
The fact that the ansatz
 depends on the integer $n$, with $ 1 \le n \le N$,
  provides a family of  quadratic algebras
 indexed by $n$. We also note that the conjugation relation (\ref{fondConj})
 shows that the quadratic algebras defined in  table (\ref{algebra})
 provide representations for the stationary state
  of the $N$-TASEP: this
 is a much more concise (albeit more abstract)  proof than the one given 
 in \cite{PEM}.

\subsection{Sector specificity} 

 We consider a basic sector  $\frs=\{s_1,\dots,s_N\}$ and let 
 $C(\frs)$ be the set of all configurations of  $\frs$.
We define
 ${\psi}{}^{(N,n)}_{\frs,\frs\setminus\{s_{n'}\}}$:
 $ V_{\frs\setminus\{s_{n'}\}}  \to  V_{\frs} $ by
\begin{eqnarray}
  \langle j_1\cdots j_L|\psi^{(N,n)}_{\frs,\frs\setminus\{s_{n'}\}} 
   |k_1\cdots k_L\rangle
=\Tr\left( a^{(N,n)}_{j_1k_1}\cdots a^{(N,n)}_{j_Lk_L}\right)
\end{eqnarray}
for $1\le n,n'\le N$,
 $ j_1\cdots j_L \in C(\frs)$
  and $ k_1\cdots k_L \in C\left(\frs\setminus\{s_{n'}\}\right)$.
In fact,   ${\psi}{}^{(N,n)}_{\frs,\frs\setminus\{s_{n'}\}}$
  is nothing but a sub-matrix of $\Psi^{(N,n)}$.
Noting  that ${M}_\frs V_{\frs}\subset V_{\frs}$ and 
${M}_{\frs\setminus\{s_{n'}\}}V_{\frs\setminus\{s_{n'}\}}
  \subset V_{\frs\setminus\{s_{n'}\}}$, we deduce from equation (\ref{fondConj})
 that 
\begin{eqnarray}
      M_\frs \psi^{(N,n)}_{\frs,\frs\setminus\{s_{n'}\} }
    = \psi^{(N,n)}_{\frs,\frs\setminus\{s_{n'}\} }
      M_{\frs\setminus\{s_{n'}\} },
\label{ConjRelNN}
\end{eqnarray}
  which is the conjugation relation between $\frs$ and $\frs\setminus\{s_{n'}\}$.
  The following property  implies that
   the conjugation matrix $\psi^{(N,n)}_{\frs,\frs\setminus\{s_{n'}\}}$
 vanishes unless  $n'=  n$. More precisely, we have 
\begin{itemize}
\item[(i)]
If $k_1\cdots k_L $ ($1\le k_\ell\le N$)
   does {\it  not} belong to $C\left(\frs\setminus\{s_n\}\right)$, then
  \begin{eqnarray}
    \Tr\left[ a^{(N,n)}_{j_1k_1}\cdots a^{(N,n)}_{j_Lk_L} \right]=0.
\label{Pte1}
  \end{eqnarray}
 \item[(ii)]  Equivalently, \begin{eqnarray} 
 \hbox{ if } \,\,\, 
  \Tr\left[ a^{(N,n)}_{j_1k_1}\cdots a^{(N,n)}_{j_Lk_L} \right]\neq0, \,\,  
 \hbox{then }  k_1\cdots k_L  \in 
   C\left(\frs\setminus\{s_n\}\right)  . \label{Pte2} \end{eqnarray}
\end{itemize}
This statement will be proved in appendix B.
Note that the sort sequences 
  of the sectors $\frs$ and $\frs\setminus\{s_n\}$ can be represented as
\begin{eqnarray}
\fl \quad
   \label{sortt}
  \overbrace{1\cdots1}^{s_1}\,2\cdots n-1\overbrace{n\cdots n}^{s_n-s_{n-1}}
  \overbrace{n+1\cdots n+1}^{s_{n+1}-s_n}\, n+2
    \cdots \quad N \quad\! \overbrace{N+1\cdots N+1}^{L-s_n}\ , \\
\fl \quad
  \underbrace{1\cdots1}_{s_1}\,2\cdots n-1 \ 
 \underbrace{n\cdots n\quad n\quad\cdots\quad n\quad\!\!}_{s_{n+1}-s_{n-1}}\, n+1
  \cdots N-1\underbrace{\quad N\quad\cdots\quad N\quad }_{L-s_n}\ ,
    \label{sort-n}
\end{eqnarray}
respectively.
The index $n$  specifies which kind of particles splits
  when the sector $\frs\setminus\{s_n\}$ is lifted to $\frs$.
We write simply
\begin{eqnarray}\label{omit-n}
   \psi_{\frs,\frs\setminus\{s_{n}\} }
   =\psi^{(N,n)}_{\frs,\frs\setminus\{s_{n}\} }.
\end{eqnarray}
We recall that $n$ appears
 explicitly in (\ref{defofhat}) for  the hat matrix  $\widehat{a}^{(N,n)}$, 
  leading to  the quadratic algebra
  generated  by the matrix elements of  ${a}^{(N,n)}$.
It is important
 to note that the sector specification property  depends on  the
 representation (\ref{solution}).

We emphasize that the  statements (\ref{Pte1}) or (\ref{Pte2})
  do not guarantee 
  that $\psi_{\frs,\frs\setminus\{s_n\}}$ is non-vanishing 
  (\textit{i.e.} meaningful).
However, based on exact calculations for
  small system sizes using Mathematica,
  we shall conjecture the stronger
  property that the mapping
  ${\psi}_{\frs,\frs\setminus\{s_n\}}$ is injective.

\subsection{Uniqueness.}
 Conjugation operators $T$  from $V_{\frs\setminus\{s_n\}}$ to  $V_{\frs}$
 that satisfy  $ M_\frs T    = T    M_{\frs\setminus\{s_n\}}$ are not unique.
 Indeed, because  of the spectral inclusion
  ${\rm Spec}(\frs) \supset {\rm Spec}(\frs\setminus\{s_n\}) $,
  there exist, in principle,
    at least ${\rm dim}\,V_{\frs\setminus\{s_n\}}$ such 
   conjugation  operators \cite{AM}.
We conjecture, however, that the  conjugation matrix
   $\psi_{\frs,\frs\setminus\{s_n\}} $ is unique if the following additional
 constraint is imposed.

\noindent
{\bf Uniqueness conjecture}:
The solution to $M_\frs \psi_{\frs,\frs\setminus\{s_n\}}
      =  \psi_{\frs,\frs\setminus\{s_n\}}
        M_{\frs\setminus\{s_n\}} $   is unique up to an overall  constant factor
when the following constraint is imposed: 
\begin{eqnarray}\label{zerocondition}
  \langle j_1\cdots j_L|
   \psi_{\frs,\frs\setminus\{s_n\}}| k_1\cdots k_L\rangle =0,
 \end{eqnarray}
for all configurations
   $j_1\cdots j_L\in C(\frs)$ and $k_1\cdots k_L\in C(\frs\setminus\{s_n\})$
  such that  $\exists i$ such that $ k_i+1\le j_i\le n $ or $ n+1\le
  j_i\le k_i $. 

 The  conjugation matrix $\psi_{\frs,\frs\setminus\{s_n\}}$
  constructed by using the operators given in (\ref{solution})
  satisfies this  condition (see the example (\ref{Nn=63}))
 \footnote{Note that the  uniqueness  conjecture  does {\it not} claim that
   the representation (\ref{solution})
   for the hat relation (\ref{hatrelation}) is unique.}.

\subsection{Conjugation relation for $\frs\supset\frt$}

Now we turn to the construction of the conjugation matrix
  between Markov matrices of arbitrary sectors $\frs$ and $\frt$
  such that $\frs\supset\frt$.
Let us set
   $\frs=\{s_1<\cdots<s_N\}$ and 
   $\frs\setminus\frt=\{ s_{n_1},\dots,s_{n_u} \}$.
Using the conjugation relation (\ref{ConjRelNN})
  between nearest-neighbor pairs repeatedly,
  we achieve the conjugation relation
\begin{eqnarray}
   \psi_{\frs\frt} M_\frt = M_\frs \psi_{\frs\frt},
\end{eqnarray}
where 
\begin{eqnarray}\label{psi=psicdotspsi}
  \psi_{\frs\frt}=
    \psi_{\frs\frs_1}\psi_{\frs_1\frs_2}\cdots
    \psi_{\frs_{u-2}\frs_{u-1}}\psi_{\frs_{u-1}\frt}
\end{eqnarray}
  with $\frs_x=\frs\setminus\{s_{n_1},\dots,s_{n_x}\}
    = \frt\cup\{s_{n_{x+1}},\dots,s_{n_u}\}$
    (for simplicity, we have omitted
    the superscripts in the $\psi$ mappings
    as in  equation (\ref{omit-n})).
Each $\psi_{\frs_{x-1}\frs_x}$ 
  is constructed by using $a^{(N+1-x,y)}$ with
$y=n_x-\#\{z|z<x,n_z<n_x\}.$ A priori, this 
  definition depends on the order chosen to 
 enumerate  the set   $\frs\setminus\frt$. However, we conjecture
 that  $\psi_{\frs\frt}$ is independent from the choice of
  the path in the Hasse diagram.

 \subsubsection{Commutativity.}

 The path independence in the  Hasse diagram can be summarized
 by the following statement,
which we have checked 
  for several values of $(N,n)$
  by using Mathematica.

\noindent
{\bf Commutativity  conjecture}:
For any  sector $\frs=\{s_1<\cdots<s_N\}$
  and  for $1\le x<y\le N$,
\begin{eqnarray}
  \psi_{\frs,\frs\setminus\{s_x\}}
   \psi_{\frs\setminus\{s_x\},\frs\setminus\{s_x,s_y\}}
  =
  \psi_{\frs,\frs\setminus\{s_y\}}
   \psi_{\frs\setminus\{s_y\},\frs\setminus\{s_x,s_y\}}.
\end{eqnarray}

\noindent
This   conjecture implies   that  
  for  two sectors $\frt \subset \frs $, 
  with $\frs\setminus\frt=\{s_{n_1},\dots, s_{n_u} \}$, 
  $\psi_{\frs\frt}$ is independent of
  the path in the Hasse diagram. In other words,  for  any reordering such that 
$\{\widetilde{n}_1,\dots, \widetilde{n}_u\}
   =\{n_1,\dots,n_u\}$,
\begin{eqnarray}
\eqalign{
  \psi_{\frs\frs_1}\psi_{\frs_1\frs_2}
  \cdots \psi_{\frs_{u-2}\frs_{u-1}}\psi_{\frs_{u-1}\frt}
 =
  \psi_{\frs\tilde{\frs}_1}\psi_{\tilde{\frs}_1\tilde{\frs}_2}
  \cdots \psi_{\tilde{\frs}_{u-2}\tilde{\frs}_{u-1}}
   \psi_{\tilde{\frs}_{u-1}\frt}
}
\end{eqnarray}
with $\frs_x=\frs\setminus\{s_{n_1},\dots,s_{n_x}\}$
  and $\tilde{\frs}_x=
   \frs\setminus\{s_{\widetilde{n}_1},\dots,s_{\widetilde{n}_x}\}$.

\noindent
{\bf Remark}:  In particular, we note that the  stationary state $|\bar P_\frs\rangle$
  of each sector $\frs=\{s_1<\cdots<s_N\}$ can be obtained  as 
\begin{eqnarray}
  |\bar P_\frs \rangle =
  \psi_{\frs\frs_1}\psi_{\frs_1\frs_2}\cdots
  \psi_{\frs_{N-2}\frs_{N-1}}\psi_{\frs_{N-1}\emptyset}|1\cdots 1\rangle ,
\end{eqnarray}
where we can chose the intermediate sectors
  $\frs_x=\{s_{n_1},\dots,s_{n_x}\}$ arbitrarily.
Since the stationary state is unique in each sector, 
  the compositions with different sets of intermediate sectors
  must be the same up to a constant factor.
This observation supports the commutativity conjecture.

\noindent
{\bf Example}: We take   $\frs=\{2,3,5,7,11,13\}$ and $\frt=\{2,7,13\}$
  so that $\frs\setminus\frt=\{3,5,11\}$.
If one chooses  $s_{n_1}=11, s_{n_2}=5$ and $s_{n_3}=3$, then 
 the intermediate conjugation matrices are constructed from 
  $ a^{(6,5)}, a^{(5,3)}$ and $a^{(4,2)}$, respectively, and this leads to 
  the conjugation matrix $\psi_{\frs\frt}$
\begin{eqnarray}
\fl\quad
  \psi_{\frs\frt}=\psi_{\{2,3,5,7,11,13\},\{2,3,5,7,13\}}
         \psi_{\{2,3,5,7,13\},\{2,3,7,13\}}
         \psi_{\{2,3,7,13\},\{2,7,13\}}. 
\end{eqnarray}
 Taking  $s_{n_1}=5, s_{n_2}=3$ and $s_{n_3}=11$,
 we obtain the following conjugation matrix:
\begin{eqnarray}
\fl\quad
  \widetilde\psi_{\frs\frt}=\psi_{\{2,3,5,7,11,13\},\{2,3,7,11,13\}}
         \psi_{\{2,3,7,11,13\},\{2,7,11,13\}}
         \psi_{\{2,7,11,13\},\{2,7,13\}},
\end{eqnarray}
where the intermediate conjugation matrices are constructed by
  $a^{(6,3)}, a^{(5,2)}$ and $a^{(4,3)}$, respectively.
The commutativity   conjecture  implies  that  $\psi_{\frs\frt} =  \widetilde\psi_{\frs\frt}.$

\subsection{Generalized Ferrari-Martin algorithm}

\begin{figure}[h]
\begin{center}
 \includegraphics[height=8cm]{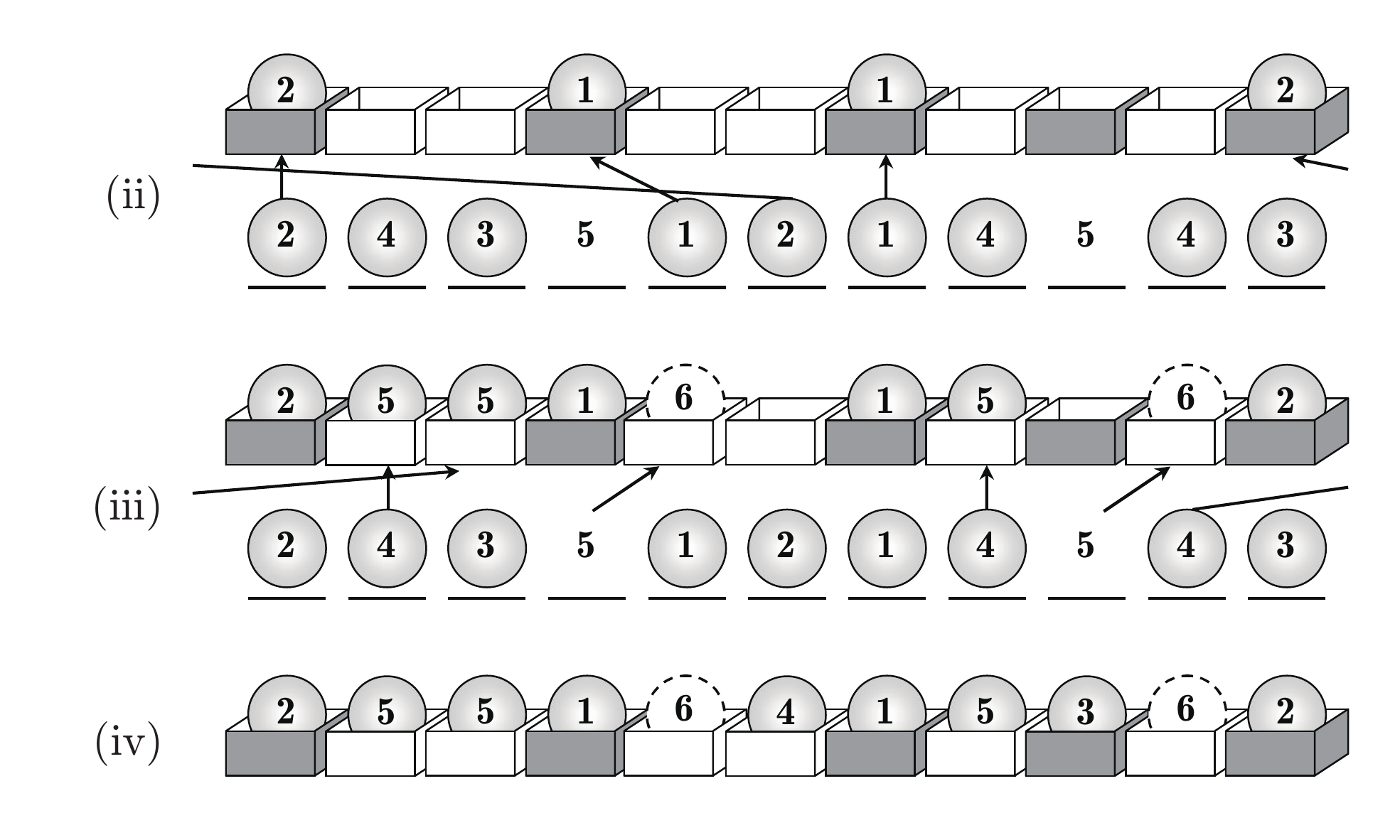}
  \caption{The generalized Ferrari-Martin algorithm that constructs
 a configuration of $N$-TASEP from
 that of $(N-1)$-TASEP.
 This figure provides the  example 
 $F(bwwbwwbwbwb$,24351214543)$
 =$25516415362.}
 \label{GFM}
\end{center}
\end{figure}

The algebraic construction that we have presented for all values
 of $n$, with $1 \le n \le N$, can be turned into an algorithm
 to calculate the stationary weights of the $N$-TASEP, that
 generalizes the original  Ferrari-Martin algorithm \cite{FM}.
Figure \ref{GFM} provides an example for $(N,n)=(5,3)$.

 Given two sectors $\frs = \{ s_1 < \dots < s_N \}$
   and $\frs\setminus\{s_n\}$, the   matrix product representation
   of the conjugation matrix  $\psi_{\frs,\frs\setminus\{s_n\}}$
 leads to  the following generalization of the Ferrari-Marin algorithm.
\begin{itemize}
\item[(i)]
Set $s_n$ black boxes and $(L-s_n)$ white boxes
  arbitrarily as $c_1\cdots c_L$ ($c_i=b,w$)
  on the upper line,
  and a configuration $k_1\cdots k_L$
  of the sector $\frs\setminus\{s_n\}$
  on the lower line.
There are
\begin{eqnarray}
\fl  \quad
     ( s_\nu-s_{\nu-1} )\ \nu\mbox{th-class particles}
     \quad (\mbox{for}\ 1\le \nu\le n-1,\ s_0=0), 
\\  \fl  \quad
     ( s_{n+1}-s_{n-1} )\ n\mbox{th-class particles},
\\ \fl   \quad
     ( s_{\nu+1}-s_\nu )\ \nu\mbox{th-class particles}
     \quad (\mbox{for}\ n+1\le \nu\le N,\ s_{N+1}=L)
\end{eqnarray}
 on the lower line.
\item[] \ As in the original Ferrari-Martin algorithm,
put particles of $\nu$th class
from $\nu=1$ to $\nu=n-1$,
according to the rule (ii-$\nu$).
\item[(ii-$\nu$)]
There are $(s_n-s_{\nu-1})$ unoccupied black boxes.
Let $\{ i_1,\dots, i_{s_\nu-s_{\nu-1}} \}$
  be the positions of the $\nu$th-class particles.
Iterate the following procedure $(s_\nu-s_{\nu-1})$ times;
find the nearest unoccupied black box $c_{i'}=b$ with
  $i'\le i_\ell$ 
   for the $\ell$th $\nu(=k_{i_\ell})$,
   and put the particle $\nu$ on it
  or on the rightmost unoccupied black box
  if  $i'$ does not exist.
\item[] \ Put particles of $(\nu+1)$th class
from $\nu=N$ to $\nu=n+1$, according to the rule (iii-$\nu$)
which is opposite to (ii).
\item[(iii-$\nu$)]
There are $(s_{\nu+1}-s_{n})$ unoccupied white boxes.
Let $\{ i_1,\dots, i_{s_{\nu+1}-s_\nu} \}$
  be the positions of the $\nu$th-class particles.
Iterate the following procedure $(s_{\nu+1}-s_\nu)$ times;
find the nearest unoccupied white box $c_{i'}=w$ with $i'\ge i_\ell$
   for the $\ell$th $\nu(=k_{i_\ell})$,
   and put the particle $\nu+1$ on it
  or on the leftmost unoccupied white box
  if  $i'$ does not exist.
(Note that for   $\nu=N$,
we ``put empty sites'' ({\it i.e.}, particles of $(N+1)$th class).)
\item[(iv)]
There are ($s_n-s_{n-1}$) unoccupied black boxes
  and ($s_{n+1}-s_n$) unoccupied white boxes.
Put $n$'s and ($n+1$)'s on them, respectively.
We have a configuration
  $F(c_1\cdots c_L, k_1\cdots k_L)$ on the upper line.
\end{itemize}

This generalized Ferrari-Martin algorithm constructs
  the same conjugation matrix:
\begin{eqnarray}\label{psi=GFM}
   \psi_{\frs,\frs\setminus\{s_n\}}=
  \sum
   |F(c_1\cdots c_L, k_1\cdots k_L)\rangle\langle k_1\cdots k_L|,
\end{eqnarray}
where $\sum$ runs over
  $ c_1\cdots c_L $ and $k_1\cdots k_L$
  with $\#\{i|c_i=b\}=s_n$ and
  $k_1\cdots k_L$ belonging to the sector $\frs\setminus\{s_n\}$.

This algorithm allows us to close the loop between this work and 
 the previous articles \cite{FM, EFM,PEM}.
In these previous works,
 the original Ferrari-Martin algorithm, in which empty sites
 played a very special role,  was used to construct a quadratic algebra
 to represent the stationary state.
  Here, we have
 found a family of matrix product representations
 that allow to split {\it any given species into two subspecies}
 (so that empty sites do not play a distinguished  role anymore).

\begin{figure}[h]
\begin{center}
\includegraphics[height=9cm]{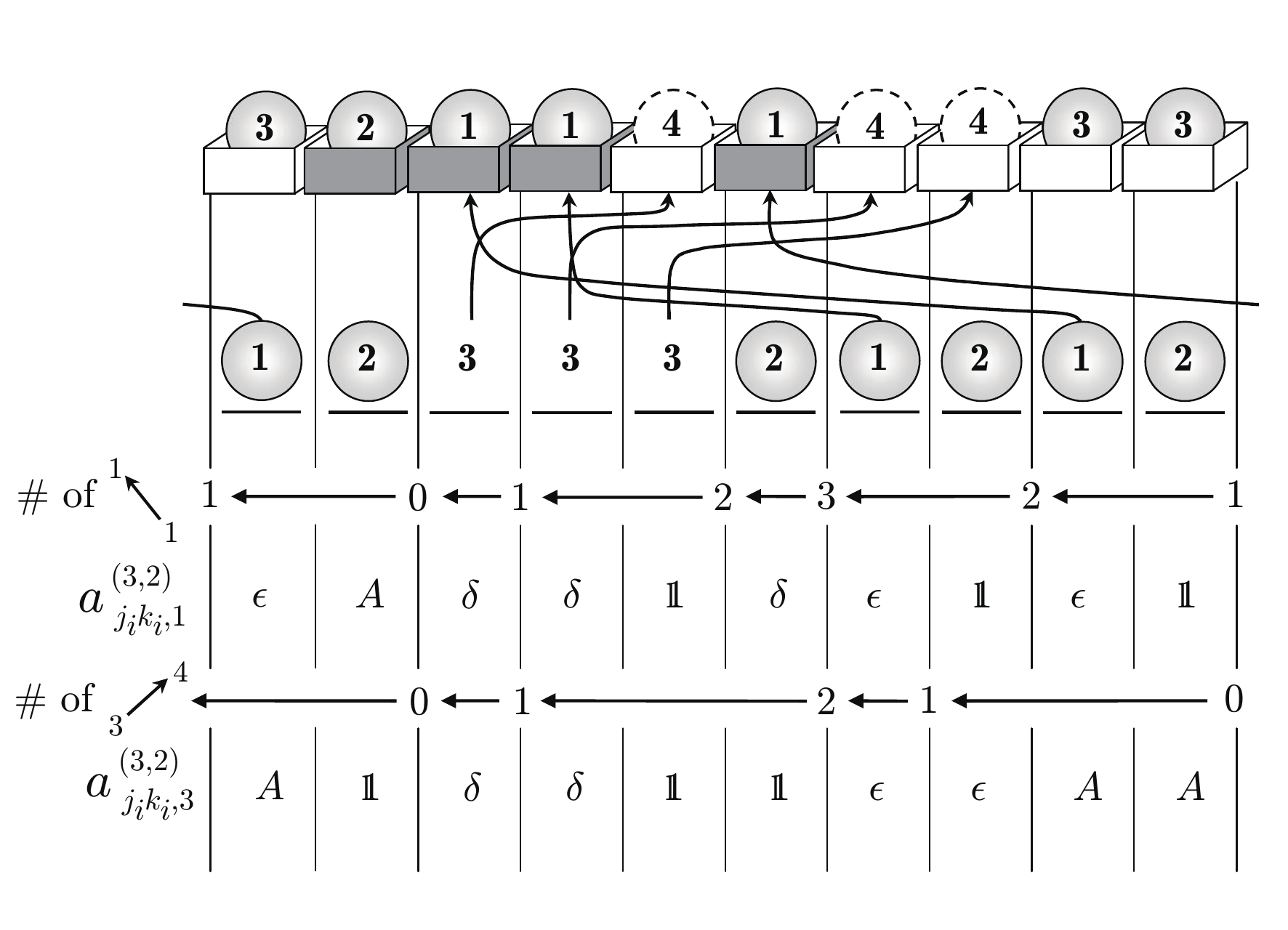}\\[-5mm]
 \caption{Correspondence between the representation (\ref{solution})
 and the generalized Ferrari-Martin algorithm.
}
\label{GMP-GFM}
\end{center}
\end{figure}

  To conclude this section,
   we precisely show  the equivalence (\ref{psi=GFM}) between
  the matrix product representation
  and the generalized Ferrari-Martin algorithm
  (GFMA).
We first note that each nonzero element of $a^{(N,n)}_{JK}$
  (\ref{solution}) has the form 
\begin{eqnarray}\label{a=aaa}
\fl \quad
  a^{(N,n)}_{JK}=
  a^{(N,n)}_{JK,1} \ot \cdots \ot a^{(N,n)}_{JK,n-1}\ot
  a^{(N,n)}_{JK,n+1} \ot \cdots \ot a^{(N,n)}_{JK,N}
   \hbox{ with } a^{(N,n)}_{JK,\nu}\in\{ \one, A, \del, \eps \}.
\end{eqnarray}
 (Note the shift in the subscripts that
 occurs for $\nu>n$).
\newcommand{\brabra}{{\langle\!\langle}}
\newcommand{\ketket}{{\rangle\!\rangle}}
Let
\begin{eqnarray}
\begin{array}{ccc}
  \A =\displaystyle\bigoplus_{\mu\ge0}\C|\mu\ketket ,\ 
  \brabra\mu | =
 (0,&\!\!\!\!\!   \dots  ,0,1,0,  \dots &  \!\!\!\!\!), 
 \   |\mu \ketket = \brabra\mu |  ^{\rm T}
\\[-4.5mm]
  & \scriptstyle{(\mu  +  1){\rm th}}  &
\end{array}
\end{eqnarray}
be the space on which the matrix $a^{(N,n)}_{JK,\nu}$ acts
(thus $a^{(N,n)}_{JK}$ acts on $\A^{\ot (N-1)}$).
The fundamental matrices $A,\del$ and $\eps$
 act on $|\mu\ketket $ as
\begin{eqnarray}
\fl \quad
  A|\mu\ketket = \left\{\begin{array}{ll}
    0  &  (\mu\ge 1), \\
     |0\ketket &  (\mu=0) ,   \end{array} \right. \quad
  \del|\mu\ketket =
  \left\{\begin{array}{ll}
    |\mu-1\ketket  &  (\mu\ge 1), \\
    0   &  (\mu=0) ,
    \end{array} \right. \quad
  \eps|\mu\ketket =  |\mu+1\ketket  .
\end{eqnarray}
The form (\ref{a=aaa}) implies that
  its trace is again decomposed as
\begin{eqnarray}\label{tr-decom}
\fl
\eqalign{
 \Tr \left(a^{(N,n)}_{j_1k_1}\cdots a^{(N,n)}_{j_Lk_L}\right)
 = \left\{ \begin{array}{l}
 \displaystyle
\prod_{1\le\nu\le N \atop \nu\neq n}
   \Tr \left(a^{(N,n)}_{j_1k_1,\nu}\cdots a^{(N,n)}_{j_Lk_L,\nu}\right) \quad 
    \left(\mbox{every }a^{(N,n)}_{j_i  k_i }\neq 0 \right),  \\
  0  \quad\quad\quad \left(\mbox{at least one $a^{(N,n)}_{j_i  k_i }=0$ }\right).
  \end{array}\right.
}
\end{eqnarray}
Furthermore, we find 
\begin{eqnarray}
\fl \quad
 \prod_{1\le\nu\le N \atop \nu\neq n}
  \Tr \left(a^{(N,n)}_{j_1k_1,\nu}\cdots a^{(N,n)}_{j_Lk_L,\nu}\right)
  =  \prod_{1\le\nu\le N \atop \nu\neq n} \sum_{\mu'_\nu\ge 0}
  \brabra \mu'_\nu| a^{(N,n)}_{j_1k_1,\nu}\cdots a^{(N,n)}_{j_Lk_L,\nu}
  | \mu'_\nu\ketket  \\
\fl \quad
  = 
  \left\{\begin{array}{ll}
  \displaystyle  \prod_{1\le\nu\le N \atop \nu\neq n}
  \brabra \mu_\nu| a^{(N,n)}_{j_1k_1,\nu}\cdots a^{(N,n)}_{j_Lk_L,\nu}
  | \mu_\nu\ketket
  =    1  &  (\ast), \\
    0   &  ({\rm otherwise}) ,
    \end{array} \right.
\end{eqnarray}
The symbol $\ast$ denotes the case where
there exists $(\mu_1,\dots\mu_{n-1},\mu_{n+1},\dots,\mu_N)$
such that 
\begin{eqnarray}\fl \label{arrow-conservation} \quad
 a^{(N,n)}_{j_1k_1,\nu}\cdots a^{(N,n)}_{j_Lk_L,\nu}
  | \mu_\nu\ketket = | \mu_\nu\ketket ,\quad
 a^{(N,n)}_{j_1k_1,\nu}\cdots a^{(N,n)}_{j_Lk_L,\nu}
  | \mu'_\nu\ketket = 0 \ ( \mu'_\nu \neq \mu_\nu).
\end{eqnarray}
The set of numbers $(\mu_1,\dots\mu_{n-1},\mu_{n+1},\dots,\mu_N)$
which satisfies this condition is unique if it exists.
This uniqueness is true for basic sectors.
On the other hand,
we draw a vertical line at each bond
in the GFMA as in  figure~\ref{GMP-GFM}.
The correspondence between the algorithm and the action
 of the operators  can be understood by
  regarding $|\mu\ketket$ as the number $\mu$ of arrows
 $\nu\to\nu\ (\nu<n)$ or
$\nu\to\nu+1 \ (\nu>n) $
crossing each vertical line:
 $\delta$ decreases 
 the number $\mu$ of arrows,
 $\epsilon$ increases $\mu$,
 $A$ tests whether $\mu=0$ or not,
 and the identity operator $\one$  indeed does nothing.
Then we find that, for given configurations $j_1\cdots j_L$
and $k_1\cdots k_L$ of $N$-species and 
$(N-1)$-species sectors,
the matrix product
$a^{(N,n)}_{j_1k_1}\cdots a^{(N,n)}_{j_Lk_L}$
gives a unique pattern of arrows (if $\ast$ is satisfied).
This means that $j_1\cdots j_L$ can be obtained
by using the GFMA from $k_1\cdots k_L$:
\begin{equation}
\fl\quad
 F(c_1\cdots c_L,k_1\cdots k_L) = j_1 \cdots j_L \quad 
{\rm with} \quad 
c_i=b(j_i\le n),w(j_i\ge n+1).
\end{equation}
This relation is true because the representation
(\ref{solution}) obeys the rule of the GFMA.
(For instance, the $\nu$th element
of  $a^{(N,n)}_{j_ik_i}$  for $j_i,k_i<n$ and $\nu\le j_i-1$
is $A$, which means that
$j_i$ can be put on the $i$th site of the upper line only when 
no arrow $\nu\to\nu$ crosses the vertical line between
sites $i$ and $i+1$.)
Thus, this graphical construction shows that
the matrix product representation and 
the GFMA are euivalent.

Figure~\ref{GMP-GFM} provides
an example of the correspondence
in the case $(N,n)=(3,2)$:
for $j_1\cdots j_{10} = 3211414433$ and 
$k_1\cdots k_{10}=1233321212$,
the actions of $a^{(3,2)}_{j_ik_i,1}$ and $a^{(3,2)}_{j_ik_i,3}$
give {\it trajectories} of the numbers of arrows $1\to 1$ and $3\to 4$,
respectively.
Namely, the condition $\ast$ is satisfied for these configurations:
one can show
\begin{eqnarray}
\fl \quad
 a^{(3,2)}_{j_1k_1,1}\cdots a^{(3,2)}_{j_{10}k_{10},1} |1\ketket = |1\ketket,
 \quad
 a^{(3,2)}_{j_1k_1,1}\cdots a^{(3,2)}_{j_{10}k_{10},1} |\mu'_1\ketket =0
 \ (\mu'_1\neq 1), \\
\fl \quad
 a^{(3,2)}_{j_1k_1,3}\cdots a^{(3,2)}_{j_{10}k_{10},3} |0\ketket = |0\ketket ,
 \quad
 a^{(3,2)}_{j_1k_1,3}\cdots a^{(3,2)}_{j_{10}k_{10},3} |\mu'_3\ketket = 0 
 \ (\mu'_3\neq 0) .
 \end{eqnarray}

\section{Concluding remarks}
\label{conclusion}

The multispecies exclusion process exhibits  rich
 combinatorial properties that can be encoded in a  recursive
 structure known as the Hasse diagram. The relevance  of this diagram
 can be understood by using identification maps: if one blurs the
 difference between  particles that  belong  to two
 adjacent classes, a simpler system
 is obtained that inherits the properties
 of the original system.
 Successive identifications  allow one
 to reduce the initial multispecies system
  to the one-species system.
 In this procedure, 
 information is irreversibly lost.

In this paper, we have shown
 that  the matrix product representation, which was
 originally used as a method for representing the stationary state,
 allows one to define conjugation
 operators that relate systems with different values of $N$
 in the TASEP case.
Therefore, the matrix ansatz
 provides a method to lift information from a simple system to
 a more complex one:
  it allows one to calculate not only the stationary state 
 but some excited states as well, that describe how the system
 relaxes towards its  steady state.
  More precisely,  each link in the Hasse diagram corresponds to 
  a  lifting operator and is 
 associated to a different quadratic algebra. Hence, the  $N$-TASEP
 leads to families of algebras,  connected through
 compatibility relations. 
We believe that this
 feature is general and that the investigation  initiated here could
 be extended in the following directions: 

 (i)  The  $N$-TASEP  is an integrable model and its
 Markov matrix can be viewed one special member of a family
 of commuting transfer matrices for the  Perk-Schultz model \cite{PS,Sc}.
 One natural question is to study  if the 
 matrix product representation for such vertex models  allows one to  
  define conjugation operators.

  (ii)  We also emphasize that  we have studied here 
 the $N$-TASEP  only, where  particles hop    in one
 definite direction.  It is  natural to expect that
  our results  could be  extended  to the $N$-PASEP.
A  naive  guess  would be  to  start with the TASEP solution (\ref{solution})
 and make the following  replacements
\begin{eqnarray}\label{replace}
\delta\to\delta_q,\ \eps\to\eps_q,\ A\to A_q
\end{eqnarray}
where $\delta_q, \eps_q$ and  $A_q$ satisfy the $q$-deformed 
 quadratic relations (\ref{q-algebra}). 
 However,  this guess 
 is correct only for  $n=1$ and $n=N$ but 
  wrong for $ 1 < n <N$. 
 The basic mathematical reason is that the  $q$-deformed  quadratic
 relations (\ref{q-algebra}) 
 are not stable by tensor product unless $q=0$: 
 a different approach  seems to be  required for solving  the
 $N$-PASEP in full generality.

\ack
C Arita was supported
  by Grant-in-Aid for Young Scientists ((B) 22740106)
  and Global COE Program
  ``Education and Research Hub for Mathematics-for-Industry''.

\appendix

\section{An explicit example}
\label{example}

Here we write down the operator-valued 
rectangular matrix $a^{(N,n)}$ (\ref{solution})
for a few values of $(N,n)$.
\begin{eqnarray}
&
a^{(2,1)}=
 \bordermatrix{
                 &{}_{1}  & {}_{2}       \cr
 \scriptstyle{1} & \one  & \del     \cr
 \scriptstyle{2} & A      & 0          \cr
 \scriptstyle{3} & \eps  & \one    },\quad
a^{(2,2)}=
 \bordermatrix{
                 &{}_{1}  & {}_{2}       \cr
 \scriptstyle{1} & \one & \del    \cr
 \scriptstyle{2} & 0      & A        \cr
 \scriptstyle{3} & \eps & \one   },
\end{eqnarray}
\begin{eqnarray}\label{example-N=3}
\fl
\eqalign{
&a^{(3,1)}= \\
&\bordermatrix{
                 &{}_{1}  & {}_{2}     & {}_{3}    \cr
 \scriptstyle{1} & \one\ot\one & \del\ot\one  & \one\ot\del  \cr
 \scriptstyle{2} & A\ot A      &    0         & 0            \cr
 \scriptstyle{3} & \eps\ot A    & \one\ot A   & 0             \cr
 \scriptstyle{4} & \one\ot\eps & \del\ot\eps  & \one\ot\one  },
}
\eqalign{
&a^{(3,2)}= \\
&\bordermatrix{
                 &{}_{1}  & {}_{2}     & {}_{3}    \cr
 \scriptstyle{1} & \one\ot\one & \del\ot\one  & \del\ot\del  \cr
 \scriptstyle{2} & 0           & A \ot\one   & A\ot\del    \cr
 \scriptstyle{3} & \eps \ot A  & \one\ot A  & 0            \cr
 \scriptstyle{4} & \eps\ot\eps & \one \ot\eps & \one\ot\one  },
}
\eqalign{
&a^{(3,3)}= \\
&\bordermatrix{
                 &{}_{1}  & {}_{2}     & {}_{3}    \cr
 \scriptstyle{1} & \one\ot\one & \del\ot\eps & \del\ot\one \cr
 \scriptstyle{2} & 0           & A\ot \one   & A \ot\del   \cr
 \scriptstyle{3} & 0           & 0             & A\ot A      \cr
 \scriptstyle{4} & \eps\ot\one & \one\ot\eps & \one\ot\one  },
}
\end{eqnarray}
\newcommand{\ao}{{\ \!\!\cdot\ \!\!}}
\begin{eqnarray}
\fl \label{Nn=63}
\eqalign{
&\quad a^{(6,3)}= \\
&
\bordermatrix{
 \  & {}_1  &{}_2  &{}_3  &{}_4  &{}_5  &{}_6 \\
{}_1 &
\one\ao\one\ao\one\ao\one\ao\one 
& \del \ao\eps \ao\one\ao\one\ao\one &
 \del \ao\one\ao\one\ao\one\ao\one
& \del \ao\one\ao\del \ao\one\ao\one
& \del \ao\one\ao\one\ao\del \ao\one &
 \del \ao\one\ao\one\ao\one\ao\del
 \\
{}_2 &  0 & A\ao\one\ao\one\ao\one\ao\one
 & A\ao\del \ao\one\ao\one\ao\one
 & A\ao\del \ao\del \ao\one\ao\one
 & A\ao\del \ao\one\ao\del \ao\one
 & A\ao\del \ao\one\ao\one\ao\del  \\
{}_3 & 0 & 0 & A\ao A\ao\one\ao\one\ao\one
 & A\ao A\ao\del \ao\one\ao\one
 & A\ao A\ao\one\ao\del \ao\one
 & A\ao A\ao\one\ao\one\ao\del
 \\
{}_4 & \eps \ao\one\ao A\ao A\ao A
 & \one\ao\eps \ao A\ao A\ao A
 & \one\ao\one\ao A\ao A\ao A
 & 0 & 0 & 0 \\
{}_5 & \eps \ao\one\ao\eps \ao A\ao A
 & \one\ao\eps \ao\eps \ao A\ao A
 & \one\ao\one\ao\eps \ao A\ao A
 & \one\ao\one\ao\one\ao A\ao A
& 0 & 0 \\
{}_6 & \eps \ao\one\ao\one\ao\eps \ao A
 & \one\ao\eps \ao\one\ao\eps \ao A
 & \one\ao\one\ao\one\ao\eps \ao A
 & \one\ao\one\ao\del\ao\eps \ao A
 & \one\ao\one\ao\one\ao\one\ao A
 & 0 \\
{}_7 & \eps \ao\one\ao\one\ao\one\ao\eps 
 & \one\ao\eps \ao\one\ao\one\ao\eps 
 & \one\ao\one\ao\one\ao\one\ao\eps
 & \one\ao\one\ao\del \ao\one\ao\eps
 & \one\ao\one\ao\one\ao\del \ao\eps 
 & \one\ao\one\ao\one\ao\one\ao\one
},
}
\end{eqnarray}
where we replaced $\ot$ by $\ao$.
We also give an example for the new conjugation matrix $\psi$ as well as the identification $\varphi$ for $\{1,2,3\}$ and $\{1,3\}$ ($(N,n)=(3,2)$) with $L=4$ (the entries of the matrices equal to $0$ are replaced by $\cdot$ for better readability):
\newcommand{\f}{\!\!\!\!1}
\newcommand{\g}{\!\!\!\!\cdot}
\begin{eqnarray}
\psi_{\{1,2,3\}\{1,3\}}=\left(\ \scriptsize
\begin{array}{cccccccccccc}
 \f & \g & \g & \g & \g & \g & \g & \g & \g & \g & \g & \g \\
 \g & \f & \f & \g & \g & \g & \g & \g & \g & \g & \g & \g \\
 \f & \f & \g & \f & \f & \g & \g & \g & \g & \g & \g & \g \\
 \g & \f & \g & \g & \f & \g & \g & \g & \g & \g & \g & \g \\
 \g & \g & \f & \g & \g & \g & \g & \g & \g & \f & \g & \g \\
 \f & \g & \f & \f & \g & \f & \g & \f & \g & \f & \f & \g \\
 \g & \g & \g & \f & \g & \f & \g & \g & \g & \g & \g & \g \\
 \g & \g & \g & \g & \f & \g & \f & \f & \f & \f & \f & \f \\
 \g & \g & \g & \g & \g & \f & \f & \g & \g & \g & \g & \g \\
 \g & \g & \g & \g & \g & \g & \f & \g & \g & \g & \g & \g \\
 \g & \g & \g & \g & \g & \g & \g & \f & \f & \g & \f & \f \\
 \g & \g & \g & \g & \g & \g & \g & \g & \f & \g & \g & \f \\
 \g & \g & \g & \f & \f & \g & \g & \g & \g & \g & \g & \g \\
 \g & \g & \g & \g & \f & \g & \g & \f & \g & \g & \g & \g \\
 \f & \f & \f & \g & \g & \f & \f & \f & \f & \g & \g & \g \\
 \g & \f & \f & \g & \g & \g & \f & \g & \f & \g & \g & \g \\
 \g & \g & \g & \g & \g & \g & \g & \f & \g & \g & \g & \g \\
 \g & \g & \f & \g & \g & \g & \g & \g & \f & \g & \g & \g \\
 \g & \g & \g & \g & \g & \g & \g & \g & \g & \f & \g & \g \\
 \g & \g & \g & \f & \g & \f & \g & \g & \g & \f & \f & \g \\
 \g & \g & \g & \g & \g & \g & \g & \g & \g & \g & \f & \f \\
 \f & \g & \g & \g & \g & \g & \g & \g & \g & \g & \g & \f \\
 \g & \g & \g & \g & \g & \f & \g & \g & \g & \g & \f & \g \\
 \f & \f & \g & \f & \f & \g & \f & \g & \g & \f & \g & \f
\end{array}
\!\!\right),
\\
\fl \quad\quad
\varphi_{\{1,3\}\{1,2,3\}}=\left(\ \scriptsize
\begin{array}{cccccccccccccccccccccccc}
 \f & \g & \f & \g & \g & \g & \g & \g & \g & \g & \g & \g & \g & \g & \g & \g & \g & \g & \g & \g & \g & \g & \g & \g \\
 \g & \f & \g & \f & \g & \g & \g & \g & \g & \g & \g & \g & \g & \g & \g & \g & \g & \g & \g & \g & \g & \g & \g & \g \\
 \g & \g & \g & \g & \f & \f & \g & \g & \g & \g & \g & \g & \g & \g & \g & \g & \g & \g & \g & \g & \g & \g & \g & \g \\
 \g & \g & \g & \g & \g & \g & \f & \g & \g & \g & \g & \g & \f & \g & \g & \g & \g & \g & \g & \g & \g & \g & \g & \g \\
 \g & \g & \g & \g & \g & \g & \g & \f & \g & \g & \g & \g & \g & \f & \g & \g & \g & \g & \g & \g & \g & \g & \g & \g \\
 \g & \g & \g & \g & \g & \g & \g & \g & \f & \g & \g & \g & \g & \g & \f & \g & \g & \g & \g & \g & \g & \g & \g & \g \\
 \g & \g & \g & \g & \g & \g & \g & \g & \g & \f & \g & \g & \g & \g & \g & \f & \g & \g & \g & \g & \g & \g & \g & \g \\
 \g & \g & \g & \g & \g & \g & \g & \g & \g & \g & \f & \g & \g & \g & \g & \g & \f & \g & \g & \g & \g & \g & \g & \g \\
 \g & \g & \g & \g & \g & \g & \g & \g & \g & \g & \g & \f & \g & \g & \g & \g & \g & \f & \g & \g & \g & \g & \g & \g \\
 \g & \g & \g & \g & \g & \g & \g & \g & \g & \g & \g & \g & \g & \g & \g & \g & \g & \g & \f & \f & \g & \g & \g & \g \\
 \g & \g & \g & \g & \g & \g & \g & \g & \g & \g & \g & \g & \g & \g & \g & \g & \g & \g & \g & \g & \f & \g & \f & \g \\
 \g & \g & \g & \g & \g & \g & \g & \g & \g & \g & \g & \g & \g & \g & \g & \g & \g & \g & \g & \g & \g & \f & \g & \f
\end{array}
\!\!\right),
\end{eqnarray}
where the bases are arranged as
1234,1243,$\dots$,4321 for $\{1,2,3\}$,
and 1223,1232,$\dots$,3221 for $\{1,3\}$.
They satisfy
\begin{eqnarray}
  \varphi_{\{1,3\}\{1,2,3\}}M_{\{1,2,3\}}
  = M_{\{1,3\}}\varphi_{\{1,3\}\{1,2,3\}},\\ 
  \psi_{\{1,2,3\}\{1,3\}}M_{\{1,3\}}
  = M_{\{1,2,3\}}\psi_{\{1,2,3\}\{1,3\}}.
\end{eqnarray}
The Markov matrix $M_{\{1,3\}}$
  has eigenvalue $E=-1$,
  and we write its corresponding eigenvector $|E\rangle$:
\begin{eqnarray}
\fl \quad
\eqalign{
|E\rangle=
|1223\rangle-|1322\rangle+|2132\rangle-|2213\rangle-|2231\rangle+|2312\rangle-|3122\rangle+|3221\rangle
}
\end{eqnarray}
The conjugation matrix $\psi_{\{1,2,3\}\{1,3\}}$
  lifts $|E\rangle$ to the sector $\{1,2,3\}$ as
\begin{eqnarray}
\fl \quad 
\eqalign{
 |E'\rangle :=
\psi_{\{1,2,3\}\{1,3\}}|E\rangle= \\
|1234\rangle-|1243\rangle+2 |1324\rangle+|1342\rangle-2 |1423\rangle-|1432\rangle-|2134\rangle+|2143\rangle \\
-2 |2314\rangle-|2341\rangle+2 |2413\rangle+|2431\rangle+|3124\rangle+2 |3142\rangle-|3214\rangle -2 |3241\rangle \\
+|3412\rangle-|3421\rangle-|4123\rangle
-2|4132\rangle+|4213\rangle+2 |4231\rangle-|4312\rangle+|4321\rangle
}
\end{eqnarray}
The vector $|E'\rangle$ is an eigenvector of $M_{\{1,2,3\}}$
 $\left(M_{\{1,2,3\}}|E'\rangle=E|E'\rangle\right)$.
The identification operator reconstructs the eigenvector
in the sector $\{1,3 \}$:
\begin{eqnarray}
  \varphi_{\{1,3\}\{1,2,3\}}|E'\rangle =3|E\rangle.
\end{eqnarray}

\section{Proof of the statement (\ref{Pte2})}
\label{app:Proof}

We shall use the following property:
let {\rm $b_i\in\{ \one, A, \del, \eps \}\ (1\le i\le L)$}.
\begin{eqnarray}
\label{delepslemma}
   \#\{i|b_i=\delta\}\neq
    \#\{i|b_i=\epsilon\}
\Rightarrow  {\rm Tr}
\left(b_1\cdots b_L\right)=0.
\end{eqnarray}
We also note the decomposition (\ref{tr-decom}).

We first consider the case $(N,n)=(3,2)$
  as an example.
Noting the properties (\ref{delepslemma}) and (\ref{tr-decom}),
we find the following necessary condition such that
$\Tr \left(a^{(3,2)}_{j_1k_1}\cdots a^{(3,2)}_{j_Lk_L}\right)\neq 0$:
every $a^{(3,2)}_{j_i  k_i }$ is nonzero and 
\begin{eqnarray}\label{numdel=numeps}
   \#  \{ i | a^{(3,2)}_{j_ik_i,\nu} =\delta \}
   =\# \{ i | a^{(3,2)}_{j_ik_i,\nu} =\epsilon \}
\end{eqnarray}
for $\nu=1,3$.
For given 3- and 2-species configurations
$j_1\cdots j_L$ and $k_1\cdots k_L$,
we have
\begin{eqnarray}
\label{num-j=1}
  \#\{i  |j_i  = 1 \} &=
     \#\left\{ i  \Big|a^{(3,2)}_{j_i  k_i ,1}=\del  \right\}
     +\, \#\{i  | j_i =k_i  =1  \},
\\
  \#\{i  |j_i  = 4 \} &=
     \#\left\{ i  \Big| a^{(3,2)}_{j_i  k_i ,3}=\eps \right\}
       +\, \#\{i  | j_i =k_i +1= 4 \},
\\
  \#\{i  |k_i  = 1\} &=
     \#\left\{ i  \Big|a^{(3,2)}_{j_i  k_i ,1}=\eps  \right\}
     +\, \#\{i  | j_i =k_i  =1  \},
\\
 \#\{i  |k_i  = 3\} &=
   \#\left\{ i  \Big|a^{(3,2)}_{j_i  k_i ,3}=\del  \right\}
       +\, \#\{i  | j_i -1=k_i =3  \},
\label{num-k=3}
\end{eqnarray}
assuming that every
 $a_{j_ik_i}^{(3,2)} \neq 0$.
(See the explicit form for $a^{(3,2)}$ (\ref{example-N=3}).)
The condition (\ref{numdel=numeps})
and the relations (\ref{num-j=1})-(\ref{num-k=3})
imply that
\begin{eqnarray}
   \#\{i|j_i=1\} = \#\{i|k_i=1\},\quad
   \#\{i|j_i=4\} = \#\{i|k_i=3\}.
\end{eqnarray}
This consequence means that
if $j_1\cdots j_L\in C(\{s_1,s_2,s_3\})$
and $\Tr \left(a^{(3,2)}_{j_1k_1}\cdots a^{(3,2)}_{j_Lk_L}\right)\neq 0$
then $k_1\cdots k_L \in C(\{s_1,s_3\})$.

In the GFMA,
the condition (\ref{numdel=numeps})
corresponds to the conservation of
the number of arrows 
when going around the ring (\ref{arrow-conservation}).
As in figure \ref{GMP-GFM},
for $j_1\cdots j_L=3211414433$
and $k_1\cdots k_L=1233321212$,
we actually observe
\begin{eqnarray}
  \#\{i|a^{(3,2)}_{j_ik_i,1}=\delta\}&=
  \#\{i|a^{(3,2)}_{j_ik_i,1}=\epsilon\}=3,\\
    \#\{i|a^{(3,2)}_{j_ik_i,3}=\delta\}&=
  \#\{i|a^{(3,2)}_{j_ik_i,3}=\epsilon\}=2.
\end{eqnarray}

For the general case,
a necessary condition for
$\Tr \left(a^{(N,n)}_{j_1k_1}\cdots a^{(N,n)}_{j_Lk_L}\right)\neq 0$
is that every $a^{(N,n)}_{j_i  k_i }$ is nonzero and 
\begin{eqnarray}
\fl\quad
   \#  \{ i | a^{(N,n)}_{j_ik_i,\nu} =\delta \}
   =\#  \{ i | a^{(N,n)}_{j_ik_i,\nu} =\epsilon \}
\ {\rm for} \ \forall\ \nu\in
\{1,\dots,n-1,n+1,\dots,N\}.
\end{eqnarray}
We can assume that every $a^{(N,n)}_{j_i  k_i }$ is nonzero.
From the definition (\ref{solution}), we find
\begin{eqnarray}
\fl \quad
\eqalign{
&  \#\{i  |j_i  =\nu \}= \\
& \left\{ \begin{array}{ll}
     \#\left\{ i  \Big|a^{(N,n)}_{j_i  k_i ,\nu}=\del  \right\}
     +\, \#\{i  | j_i =k_i  =\nu  \}
    &  (1\le\nu\le n-1), 
\vspace{1mm} \\
     \#\left\{ i  \Big| a^{(N,n)}_{j_i  k_i ,\nu-1}=\eps \right\}
       +\, \#\{i  | j_i =k_i +1=\nu  \}
    &  (n+2\le\nu\le N+1),
 \end{array}\right.
}
\\
\fl \quad
\eqalign{
&  \#\{i  |k_i  =\nu\}= \\
& \left\{ \begin{array}{ll}
     \#\left\{ i  \Big|a^{(N,n)}_{j_i  k_i ,\nu}=\eps  \right\}
     +\, \#\{i  | j_i =k_i  =\nu  \}
    &  (1\le\nu\le n-1), 
\vspace{1mm} \\
     \#\left\{ i  \Big|a^{(N,n)}_{j_i  k_i ,\nu}=\del  \right\}
       +\, \#\{i  | j_i -1=k_i =\nu  \}
    &  (n+1\le\nu\le N).
 \end{array}\right.
}
\end{eqnarray}
Consequently, we have
\begin{eqnarray}
  \#\{i  |j_i  =\nu\} &= \#\{i  |k_i  =\nu\}
         \quad{\rm for}\ 1\le\nu\le n-1,  \\
  \#\{i  |j_i  =\nu\} &= \#\{i  |k_i  = \nu-1\} 
       \quad{\rm for}\  n+2\le\nu\le N+1,
\end{eqnarray}
which exactly means that
if $j_1\cdots j_L\in C(\frs)\ (\frs=\{s_1,\dots,s_N\})$
and $\Tr \left(a^{(N,n)}_{j_1k_1}\cdots a^{(N,n)}_{j_Lk_L}\right)\neq 0$,
then $k_1\cdots k_L\in\frs\setminus\{s_n\}$.
(See the figures of the sort sequences (\ref{sortt}) and (\ref{sort-n})).
 We emphasize, however, that  even if  $ j_1\cdots j_L \in C(\frs)$
  and $ k_1\cdots k_L \in C\left(\frs\setminus\{s_{n}\}\right)$, the matrix element 
  $\Tr\left[ a^{(N,n)}_{j_1k_1}\cdots a^{(N,n)}_{j_Lk_L} \right]$
  can still be equal to  zero.

\end{document}